\documentclass[aps,prd,amsmath,twocolumn,notitlepage,showpacs,superscriptaddress,nofootinbib,usenatbib]{revtex4-1}
\def \be {\begin{equation}} 

\def \ee {\end{equation}} 
\def \bea {\begin{eqnarray}} 
\def \eea {\end{eqnarray}} 

\usepackage{graphicx}
\usepackage{dcolumn}
\usepackage{bm}
\usepackage{epsfig} 
\usepackage{color}

\newcommand{\BibitemShut}[1]{}

\newcommand*{\ltsim}{\ {\raise-.75ex\hbox{$\buildrel<\over\sim$}}\ }
\newcommand*{\gtsim}{\ {\raise-.75ex\hbox{$\buildrel>\over\sim$}}\ }
\newcommand*{\proptosim}{\ {\raise-.75ex\hbox{$\buildrel\propto\over\sim$}}\ }

\begin{document}
\title{Variation in the fine-structure constant, distance-duality relation and the next generation of high-resolution spectrograph}

\author{Rodrigo S. Gon\c{c}alves} \email{rsousa@on.br}
\affiliation{Observat\'orio Nacional, 20921-400, Rio de Janeiro - RJ, Brasil}
\affiliation{Departamento de F\'{\i}sica, Universidade Federal do Maranh\~ao, 65080-805, S\~ao Lu\'{\i}s, Maranh\~ao, Brasil}
\author{Susana Landau} \email{slandau@df.uba.ar}
\affiliation{CONICET, Godoy Cruz 2290, 1425 Ciudad Aut\'onoma de Buenos Aires, Argentina.} 
\affiliation{ Departamento de F\'{\i}sica and IFIBA, Facultad de Ciencias Exactas y Naturales, Universidad de Buenos Aires, Ciudad Universitaria - Pab. I, Buenos Aires 1428, Argentina.}
\author{Jailson S. Alcaniz}\email{alcaniz@on.br}
\affiliation{Observat\'orio Nacional, 20921-400, Rio de Janeiro - RJ, Brasil}
\affiliation{Departamento de F\'isica, Universidade Federal do Rio Grande do Norte, Natal, RN 59072-970, Brasil}
\author{Rodrigo F. L. Holanda}\email{holanda@fisica.ufrn.br}
\affiliation{Departamento de F\'isica, Universidade Federal do Rio Grande do Norte, Natal, RN 59072-970, Brasil}
\affiliation{Departamento de F\'{\i}sica, Universidade Federal de Sergipe, S\~ao Crist\'ov\~ao, SE 58429-900, Brasil}
\date{\today}

\begin{abstract}
The possibility of variation of the fundamental constants of nature has been a long-standing question, with important consequences for fundamental physics and cosmology. In particular, it has been shown that  variations in the fine-structure constant, $\alpha$, are directly related to violation of the distance duality relation (DDR), which holds true as long as photons travel on unique null geodesics and their number is conserved. In this paper we use the currently available measurements of ${\Delta \alpha}/{\alpha}$ to impose the most stringent constraints on departures of the DDR to date, here quantified by the parameter $\eta$.  We also perform a forecast analysis to discuss the ability of the new generation of high-resolution spectrographs, like ESPRESSO/VLT and E-ELT-HIRES, to constrain the DDR parameter $\eta$. From the current data we obtain constraints on $\eta$ of the order of $10^{-7}$ whereas the forecasted constraints are two orders of magnitude lower. Considering the expected level of uncertainties of the upcoming measurements, we also estimate the necessary number of data points to confirm the hypotheses behind the DDR.

\end{abstract}

\maketitle

\section{Introduction}

The search for space-time dependence of the fundamental constants is crucial to build empirical grounds of our current physical theories and also to explore signs of new physics that might manifest through small deviations~\cite{dirac1} (see also ~\cite{Uzan,Martins2017} for recent reviews).  In this regard, several experiments and observations have tested whether or not the  fundamental constants of physics are indeed constants, being grouped into astronomical and local methods. The latter ones include geophysical methods such as that for the natural nuclear reactor that operated about $1.8 \times 10^9$ yrs ago in Oklo, Gabon \cite{1996NuPhB.480...37D,Petrov06,Gould06}, the analysis of natural long-lived $\beta$ decayers in geological minerals and meteorites \cite{dyson, sisterna, Olive04b}, and laboratory measurements such as comparisons of rates between several clocks with different atomic numbers \cite{pres,Peik04,rosenband08}. The astronomical methods are based mainly on the analysis of spectra from high-redshift quasar absorption systems \cite{bahc, lev, murphy1, murphy2, Webb99, webb2}. Besides, further constraints on a possible variation of the fine structure constant ($\alpha$) can be obtained comparing X-ray and SZ measurements in galaxy clusters \cite{Galli13,Holanda16,Holanda16b,Martino16,Martino16b,Colaco:2019fvl}. The variation of the fundamental constants in  the early universe can also be analyzed from primordial nucleosynthesis predictions of the abundance of the light elements \cite{bergstrom,mosquera} and the Cosmic Microwave Background (CMB) fluctuation spectrum \cite{bat,avelino,Planck2015,obryan15} -- see e.g. \cite{Uzan,Yuri} for  extensive discussions of the many observational  techniques.

From the theoretical point of view, the attempt to unify the four fundamental interactions resulted in the development of multidimensional theories  \cite{Pablo,Uzan,Yuri} like string-motivated field theories, related brane-world theories, and (related or not) Kaluza-Klein theories, which predict at the low energy limit a dependence of the fundamental constants with time and space. Later on, phenomenological models which specifically  study a potential variation of one fundamental constant were also proposed \citep{bekenstein82,bekenstein2002,bsm02,BM05}. Moreover, among all tests of the Einstein Equivalence Principle (EEP) -- a cornerstone of the General Relativity (GR) -- the search for spatial and temporal variations of the fundamental constants is certainly an important way to test Local Position Invariance. Most of the theories that predict a time variation of the fine structure constant incorporate a non-minimal multiplicative coupling between the scalar field responsible for the variation and the matter fields. This coupling implies a non-conservation of the photon number along geodesics which has several observational consequences. For instance, Refs. \cite{Hees2014,Minazzoli2014} showed that variations of the fine structure constant and violations of the distance-duality relation (DDR)
\begin{equation}
 \frac{d_L}{d_A(1+z)^2} = \eta\;,
\label{dr}
\end{equation}
with $\eta = 1$, are intimately and unequivocally related, where $d_L$ and $d_A$ are the luminosity and angular diameter distances, respectively. As is well known, the DDR is valid as long as photon number is conserved and gravity is described by a metric theory with photons traveling on unique null geodesics -- see e.g. \cite{2007GReGr..39.1047E,Bassett:2003vu,Goncalves:2011ha,Holanda:2012at} (see also \cite{Santana:2017zvy}). Recently, a number of analysis have tested the DDR with both cosmological and local data (see e.g. \cite{Ellis:2013cu,Ruan:2018,Lin:2018,Ma:2018,daCosta:2015,Holanda:2012at} and references therein).

It is well true that the results of most experiments and observations do not show evidence for variations of the fundamental constants. In 1999, however, Webb et al \cite{Webb99}  claimed a detection of a variation in $\alpha$ from observations made with the Keck telescope. Nevertheless, an independent analysis performed
with UVES at the Very Large Telescope (VLT) some years later provided null results \cite{srianand}. Contrary to the previous results, another analysis using VLT/UVES data  also suggested a variation in $\alpha$, but now with $\alpha$ increasing with redshift \cite{webb11,King12}.  However, more recently, a recalculation of systematic errors using new techniques showed  that there is no compelling evidence for any variation in $\alpha$ from quasar data \cite{whit}. Finally, a  program using the world's  three largest optical telescopes (VLT, Keck and Subaru) was developed specifically to test  the stability of fundamental couplings. No evidence for variation in $\alpha$ or the proton to electron mass ratio,   $\mu = {m_p}/{m_e}$, was found \cite{Molaro2013,Rahmani2013,Evans2014}.  

It is worth noticing that the above measurements  
were performed with  spectrograph such as UVES, HARPS or Keck-HIRES, which  are far from optimal for testing possible variations of fundamental constants. Therefore,  more precise measurements using the new generation of high-resolution spectrograph, like ESPRESSO for the VLT \cite{ESPRESSO} and E-ELT-HIRES for the E-ELT \cite{ELT}, are expected to significantly improve the precision of the data and, crucially, have a much better control over possible systematics. 

The goal of this paper is twofold. First, to discuss constraints on the DDR from the best currently available measurements of variations of the fine structure constant ${\Delta \alpha}/{\alpha}$. Second, to perform a forecast analysis to discuss the ability of the next generation of high-resolution spectrograph to constrain departures from the DDR. In order to perform our analyses, we assume different theoretical parameterizations for $\eta$ and show that the results obtained are independent of them. This paper is organized as follows. In Section \ref{obs}, we introduce the data set and the possible targets of future missions that will be used for the statistical and forecast  analyses. We also present the $\eta$ parameterizations assumed in the paper and the first constraints on $\eta$  from the current observational data. Section \ref{stats} describes the statistical and forecast analyses performed considering the uncertainty expectations of the missions ESPRESSO/VLT and E-ELT/HIRES as well as a discussion of the results. We end the paper by summarizing the main results in section \ref{conclusions}.
 
 \begin{table}[t]
\centering
\begin{tabular}{|c|c|c|c|}
\hline
Object \,\,& \,\,$z$ \,\,& \,\, ${\Delta \alpha}/{\alpha}$ [$\times 10^{-6}$]  \,\,& \,\, Reference\\
\hline
J0026−2857 	\,\,& \,\, 1.02	\,\,& \,\,  $ 3.5 \pm	 8.9$  \,\,& \,\, \cite{Murphy2016}\\
J0058+0041 \,\,& \,\,1.07	\,\,& \,\,  $-1.4 \pm	 7.2$  \,\,& \,\,  \cite{Murphy2016}\\
3 sources\,\,& \,\,1.08	\,\,& \,\,  $ 4.3 \pm	 3.4$ \,\,& \,\,  \cite{Songaila2014} \\
HS1549+1919\,\,& \,\,1.14	\,\,& \,\,  $-7.5 \pm	 5.5$  \,\,& \,\,  \cite{Evans2014}\\
HE0515−4414\,\,& \,\,1.15	\,\,& \,\,  $-1.4 \pm	 0.9$ \,\,& \,\,  \cite{Kotus2017} \\
J1237+0106\,\,& \,\,1.31	\,\,& \,\,  $-4.5 \pm	 8.7$  \,\,& \,\,\cite{Murphy2016}\\
HS1549+1919\,\,& \,\,1.34	\,\,& \,\,  $-0.7 \pm	 6.6$  \,\,& \,\,  \cite{Evans2014} \\
J0841+0312\,\,& \,\,1.34	\,\,& \,\,  $ 3.0 \pm	 4.0$  \,\,& \,\,  \cite{Murphy2016}\\
J0841+0312\,\,& \,\,1.34	\,\,& \,\,  $ 5.7 \pm	 4.7$   \,\,& \,\ \cite{Murphy2016}\\
J0108−0037\,\,& \,\,1.37	\,\,& \,\,  $-8.4 \pm	 7.3$  \,\,& \,\,  \cite{Murphy2016}\\
HE0001−2340\,\,& \,\,1.58	\,\,& \,\,  $-1.5 \pm	 2.6$  \,\,& \,\,  \cite{Agafonova2011}\\
J1029+1039\,\,& \,\,1.62	\,\,& \,\,  $-1.7 \pm	10.1$  \,\,& \,\,  \cite{Murphy2016}\\
HE1104−1805\,\,& \,\,1.66	\,\,& \,\,  $-4.7 \pm	 5.3$  \,\,& \,\,  \cite{Songaila2014}\\
HE2217−2818\,\,& \,\,1.69	\,\,& \,\,  $ 1.3 \pm	 2.6$   \,\,& \,\,  \cite{Molaro2013}\\
HS1946+7658\,\,& \,\,1.74	\,\,& \,\,  $-7.9 \pm	 6.2$  \,\,& \,\,  \cite{Songaila2014}  \\
HS1549+1919\,\,& \,\,1.80	\,\,& \,\,  $-6.4 \pm	 7.2$  \,\,& \,\,   \cite{Evans2014} \\
Q1103−2645\,\,& \,\,1.84	\,\,& \,\,  $ 3.5 \pm	 2.5$  \,\,& \,\, \cite{Bainbridge2017}\\
Q2206−1958\,\,& \,\,1.92	\,\,& \,\,  $-4.6 \pm	 6.4$  \,\,& \,\,  \cite{Murphy2016}\\
Q1755+57\,\,& \,\,1.97	\,\,& \,\,  $ 4.7 \pm	 4.7$  \,\,& \,\,  \cite{Murphy2016}\\
PHL957\,\,& \,\,2.31	\,\,& \,\,  $-0.7 \pm	 6.8$  \,\,& \,\,  \cite{Murphy2016}\\
PHL957\,\,& \,\,2.31	\,\,& \,\,  $-0.2 \pm	12.9$  \,\,& \,\,  \cite{Murphy2016}\\
\hline
\end{tabular}
\caption{Current measurements of $\frac{\Delta \alpha}{\alpha}$ from the ESO UVES Large Program. The first column shows the source along the line of sight, the redshift of the absorption system is shown in the second column, while the reported value of $\frac{\Delta \alpha}{\alpha}$ and the corresponding $1\sigma$ error in units of $10^{-6}$ are shown in the third column; the last column shows the respective reference.} 
\label{DS1}
\end{table}

\begin{figure*}
\centering
\includegraphics[width=0.49\textwidth]{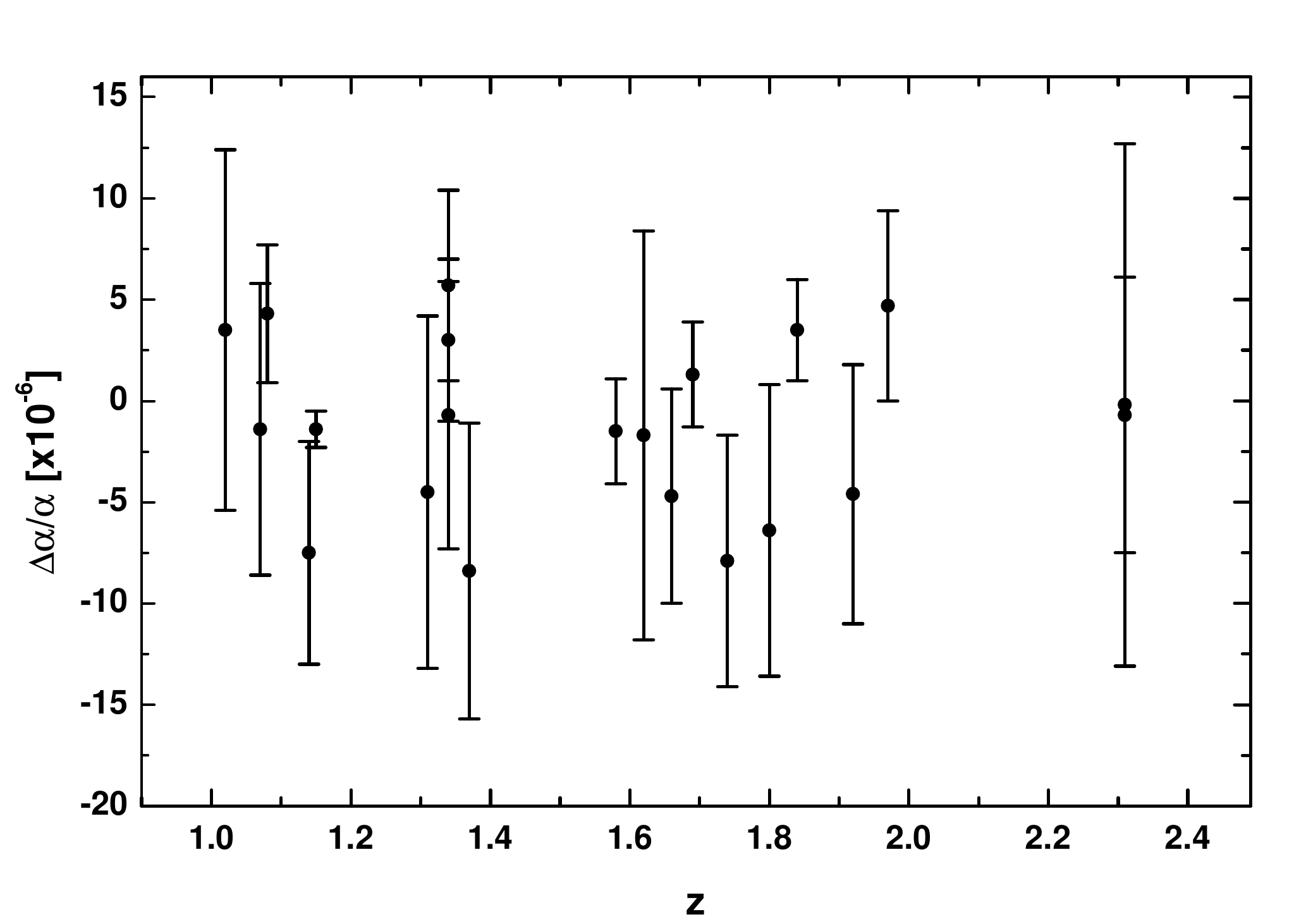}  
\includegraphics[width=0.49\textwidth]{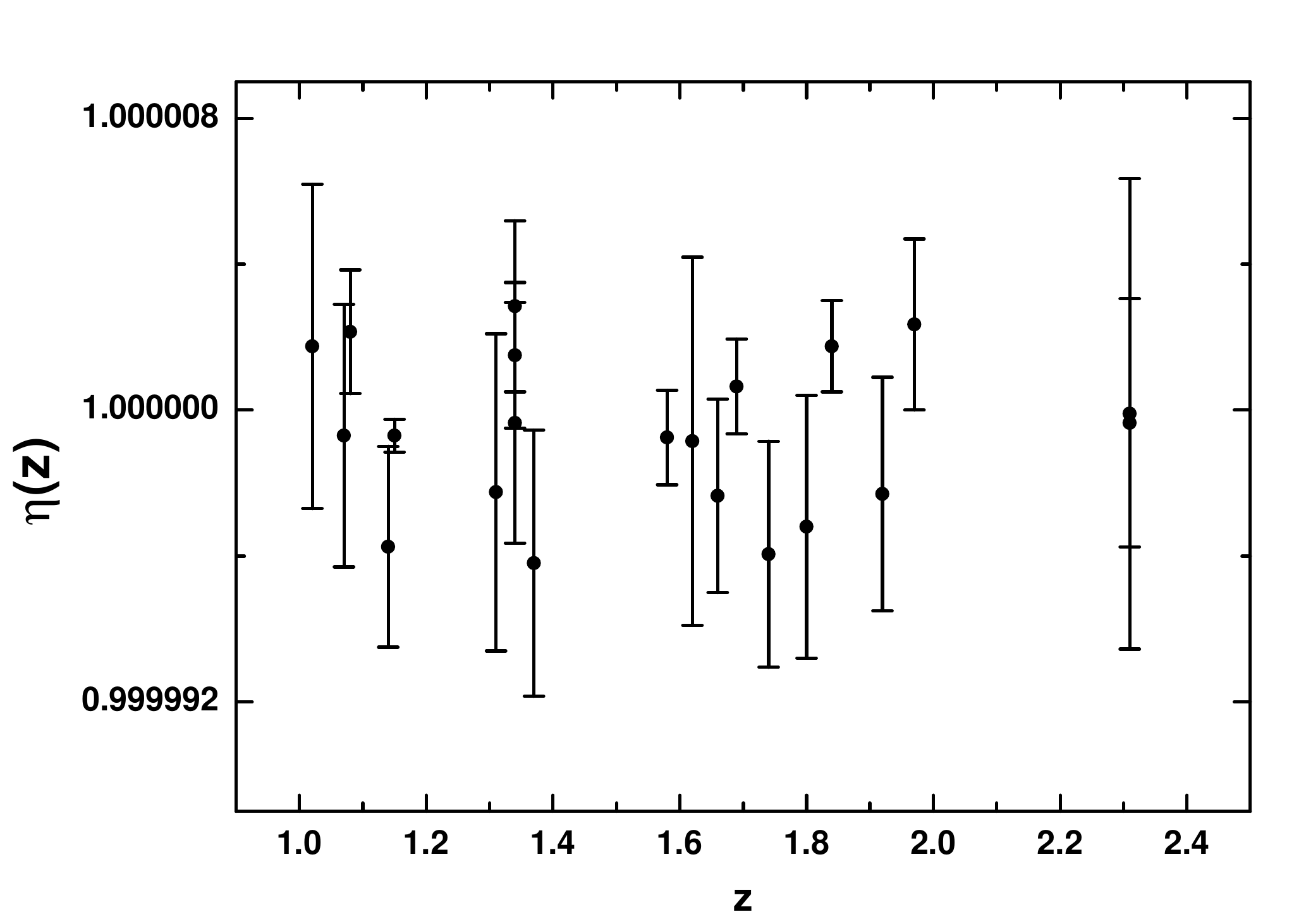} 
\caption{{\it{Left)}} Measurements of ${\Delta \alpha}/{\alpha}$ taken from Table \ref{DS1}. {\it{Right)}} Inferred values of DDR parameter $\eta(z)$ from the observational data of Table \ref{DS1} using P1.}
\label{ObsAlpha}
\end{figure*}

\section{Data Set and Constraints}
\label{obs}

\subsection{Data sets}

In the previous section we mentioned the results reported by different groups regarding measurements of a possible variation of $\alpha$ from astronomical observations. A specific concern regarding the results of Refs. \cite{Webb99,srianand,webb11,King12} is that they are based on archival data, that is, the data were not taken for the specific purpose of measuring  variation of fundamental constants. Furthermore, the data acquisition procedures were also far from ideal, particularly regarding the key issue of wavelength calibration. 

The ESO UVES Large Program was developed to confirm these results being the only large program dedicated specifically to test possible variations of fundamental constants, with an optimized sample and methodology. The results obtained for the variation of $\alpha$ are shown in Table \ref{DS1} and will be used in this paper to derive constraints on the $\eta$ parameter of Eq. (\ref{dr}) as well as to perform a forecast analysis to estimate the number of measurements needed to verify the validity  of the DDR (see Section \ref{stats}). 

On the  other hand, Leite et al.  \cite{Leite2016}  identified  a list of  14 $\alpha$  targets in the redshift range $1.35 < z < 3.02$ to be observed during the ESPRESSO Fundamental Physics Guaranteed Time Observations (GTO). Table \ref{DS2} shows the best currently available measurements of $\alpha$, among the targets accessible to ESPRESSO. Throughout this paper we refer to this data set as DS2, and will use them to produce simulated data both for the ESPRESSO/VLT and E-ELT/HIRES future missions. For the uncertainties, we follow the analysis of Leite et al \cite{Leite2016} and consider two scenarios: baseline and ideal which intend to  bracket the expected performance of both missions \cite{Martins2017}.  For the ESPRESSO/VLT we consider  $\sigma^{\rm baseline}_{\rm{VLT}} = 6 \times 10^{-7}$ and $\sigma^{\rm ideal}_{\rm{VLT}} = 2 \times 10^{-7}$ whereas for the E-ELT-HIRES, $\sigma^{\rm baseline}_{\rm{ELT}} = 1 \times 10^{-7}$ and $\sigma^{\rm ideal}_{\rm{ELT}} = 0.3 \times 10^{-7}$.

\begin{table}
\centering
\begin{tabular}{|c|c|c|c|}
\hline
Object \,\,& \,\, $z$ \,\,& \,\, ${\Delta \alpha}/{\alpha}$ [$\times 10^{-6}$] \,\,& \,\, Reference \\
\hline
J0350−3811 \,\,& \,\,  1.35 	\,\,& \,\,  $ -4.0 \pm	 2.3$ \,\,& \,\, \cite{Murphyphd}\\
J0407−4410 \,\,& \,\, 1.43 	\,\,& \,\,  $-21.3 \pm	 3.6$ \,\,& \,\,  \cite{Kingphd} \\
J0431−4855 \,\,& \,\, 1.69 	\,\,& \,\,  $  1.3 \pm	 2.4$  \,\,& \,\ \cite{Kingphd} \\
J0530−2503 \,\,& \,\  1.77 	\,\,& \,\,  $  8.4 \pm	 4.4$ \,\,& \,\ \cite{Kingphd}  \\
J1103−2645 \,\,& \,\  1.84 	\,\,& \,\,  $  5.6 \pm	 2.6$ \,\,& \,\ \cite{Bainbridge2017,Molaro2008}\\
J1159+0112 \,\,& \,\  1.86 	\,\,& \,\,  $ -9.9 \pm	 4.9$ \,\,& \,\ \cite{Kingphd}\\
J1334+1649 \,\,& \,\  1.92 	\,\,& \,\,  $  8.5 \pm	 3.8$ \,\,& \,\  \cite{Kingphd}\\
HE1347−2457 \,\,& \,\  1.94 	\,\,& \,\,  $  5.1 \pm	 4.4$ \,\,& \,\ \cite{Molaro2008}\\
J2209−1944 \,\,& \,\  2.14 	\,\,& \,\,  $  6.7 \pm	 3.5$ \,\,& \,\ \cite{Murphy16,Kingphd}\\
HE2217−2818 \,\,& \,\  2.15 	\,\,& \,\,  $  5.2 \pm	 4.3$ \,\,& \,\  \cite{Molaro2013}\\
Q2230+0232 \,\,& \,\  2.28 	\,\,& \,\,  $  7.5 \pm	 3.7$ \,\,& \,\ \cite{Murphyphd} \\
J2335−0908 \,\,& \,\  2.43 	\,\,& \,\,  $-12.2 \pm	 3.8$ \,\,& \,\  \cite{Kingphd} \\
J2335−0908 \,\,& \,\  2.59 	\,\,& \,\,  $  5.7 \pm	 3.4$ \,\,& \,\  \cite{Kingphd} \\
Q2343+1232 \,\,& \,\  3.02 	\,\,& \,\,  $-27.9 \pm	34.2$  \,\,& \,\ \cite{Murphyphd}\\
\hline
\end{tabular}
\caption{Possible targets of  ESPRESSO according to the best currently available measurements of ${\Delta \alpha}/{\alpha}$. {The first column shows the quasar name, the redshift of the absorption system is shown in the second column, while the reported value of $\frac{\Delta \alpha}{\alpha}$ and the corresponding $1\sigma$ error in units of $10^{-6}$ are shown in the third column; the last column shows the respective reference.}}
\label{DS2}
\end{table}

\subsection{Theoretical expressions for $\eta(z)$} 

\label{param}
As mentioned earlier,  Refs. \cite{Hees2014,Minazzoli2014} derived a direct relation between variations of $\alpha$ and violation of the DDR, i.e.,
 \begin{equation}
\label{FromAlphaToEta}
\frac{\Delta \alpha}{\alpha} = \eta^2(z) - 1  \rightarrow \eta(z) = \sqrt{\frac{\Delta \alpha}{\alpha} + 1} \;.
\end{equation}
Considering the above expression, we show in the left panel of Fig. \ref{ObsAlpha}  the original values of ${\Delta \alpha}/{\alpha}$ (displayed in Table \ref{DS1}) whereas the corresponding values of $\eta(z)$ are shown in the right panel of the same figure.  

In order to constrain a possible violation of the DDR, we explore three parameterizations of $\eta(z)$: 
\begin{itemize}

\item P1: $\eta(z) = 1 + \eta_{0}   z$\;,

\item P2: $\eta(z) = 1 + \eta_0  \left(\frac{z}{1+z}\right)$\;,
          
\item P3: $\eta(z) =  (1 + z)^{\eta_0}$\;,

\end{itemize}
which cover a wide range of possibilities, as discussed in ~\cite{Holanda:2012at}. The first one is a natural approximation to the problem but it diverges at large redshifts whereas the second parameterization fixes the divergence problem. The third parameterization accounts for a general photon attenuation, and  was introduced originally in the context of a departure from CMB's transparency~\cite{AVJ2009}. As shown in Section \ref{stats}, our results are independent of the parameterization adopted.

\begin{table}
\centering
\begin{tabular}{|c|c|c|}
\hline
Parameterization \,\,& \,\,  $\eta_0 \pm 1\sigma$ [$\times 10^{-7}$] \,\, \\ 
\hline
P1  \,\,& $ {-1.6}^{+2.5}_{-2.4}  $\,\, \\ 
P2  \,\,&  ${-4.6}^{+6.5}_{-6.2}  $\,\, \\ 
P3  \,\,&  ${-2.9}^{+4.0}_{-3.9}  $\,\, \\ 
\hline
\end{tabular}
\caption{Constraints on the DDR parameter $\eta_0$ from the current measurements of  ${\Delta \alpha}/{\alpha}$ displayed in Table \ref{DS1}.}
\label{TableIII}
\end{table}

\begin{figure*}
\centering
\includegraphics[width=0.49\textwidth]{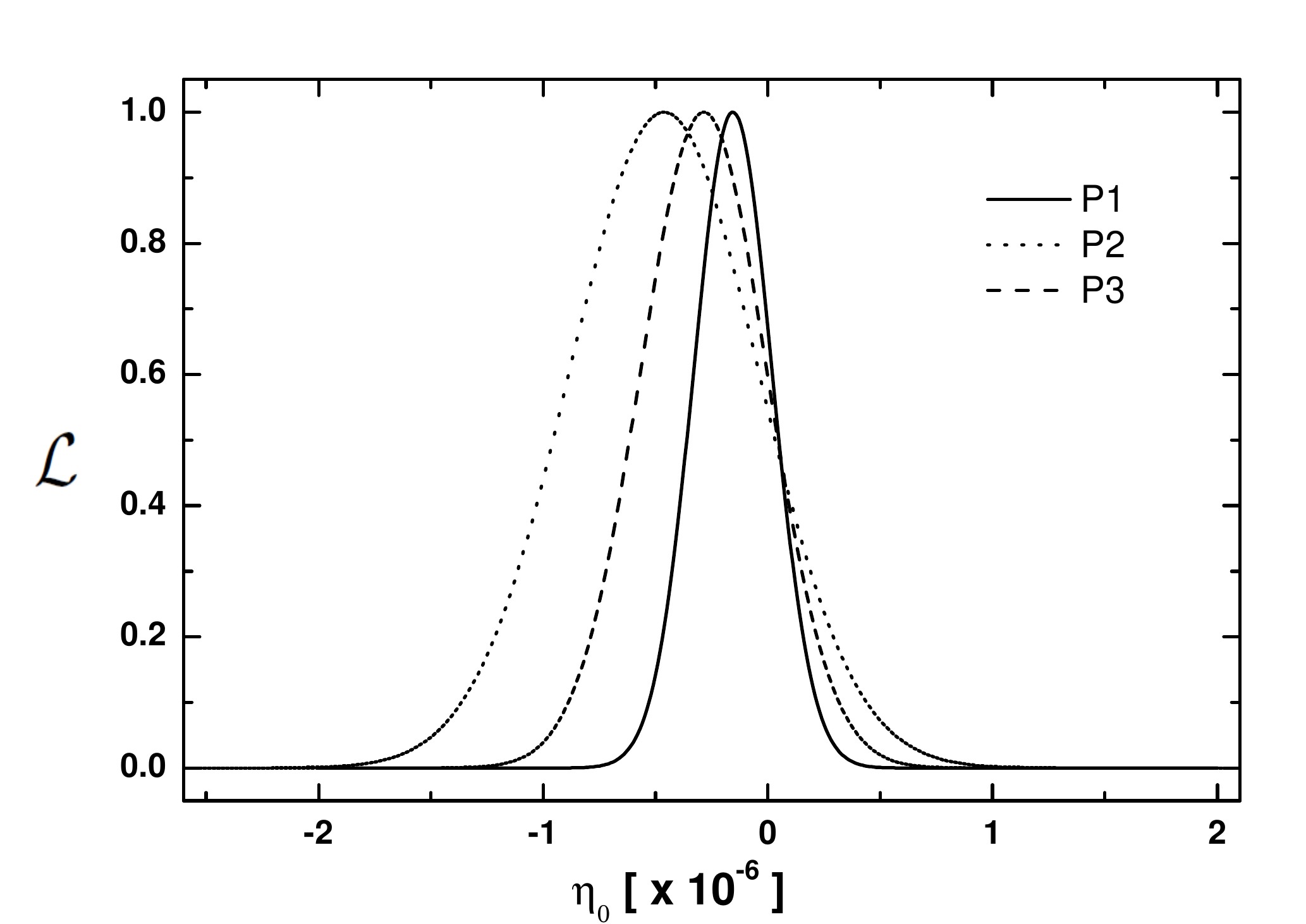}  
\includegraphics[width=0.49\textwidth]{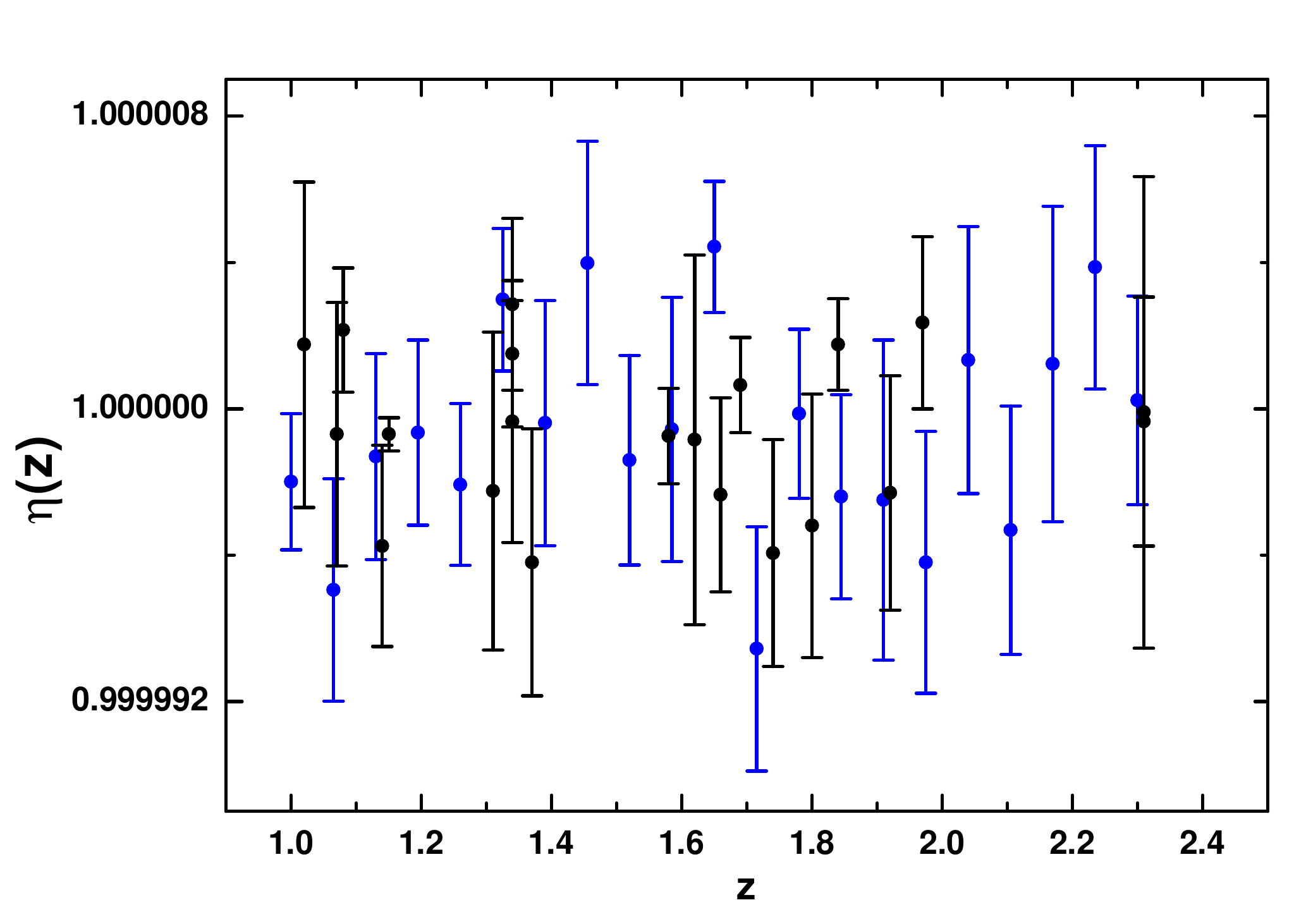} 
\caption{{\it{Left)}} Likelihood of the DDR parameter $\eta_0$ for the different parameterizations discussed in the text. {\it{Right)}} Comparision between the  inferred values of DDR parameter $\eta(z)$ from the observational data of Table \ref{DS1} (black points) and one realisation of a simulated data set (blue points).}
\label{Chi2Pars}
\end{figure*} 

\subsection{Current Constraints}

We obtain the best fit values and errors for the parameter $\eta_0$ through a $\chi^{2}$ statistics using the data set shown in Table \ref{DS1}. The corresponding likelihood, ${\cal{L}} \propto e^{- \chi^2}$,  as a function of $\eta_0$ is shown in Fig.~\ref{Chi2Pars} (left) whereas the numerical results are displayed in Table \ref{TableIII}. For all parameterizations, both the  values are of the order of $10^{-7}$, which essentially means that deviations from the DDR obtained from the current measurements of  ${\Delta \alpha}/{\alpha}$ are very small, with  $\eta(z) \sim 1$. It is worth mentioning that when compared with recent constraints on the DDR parameter from cosmological observations (see e.g. \cite{Ellis:2013cu,Ruan:2018,Lin:2018,Lv:2016,Ma:2018,daCosta:2015,Holanda:2012at}) the above results are orders of magnitude more restrictive.

We also compare our results with the ones reported by \cite{Hees2014} using measurements of $\Delta \alpha/\alpha$ from 154 absorbers observed with VLT and 128 absorbers observed at the Keck observatory. The former (latter) measurements provide ($\times 10^{-7}$): $\eta_0=8.4 \pm 3.5 $ ($\eta_0=-16 \pm 6 $), $\eta_0=20 \pm 10 $ ($\eta_0=-49 \pm 17$) and $\eta_0=14 \pm 6 $ ($\eta_0=-30 \pm 11 $ )  for P1, P2 and P3 parameterizations, respectively. Clearly, the two data sets show incompatible results which were interpreted  by the authors as due to a possible variation of $\alpha$ in the Northern and Southern hemispheres.


\begin{figure*}
\centering
\includegraphics[width=0.329\textwidth]{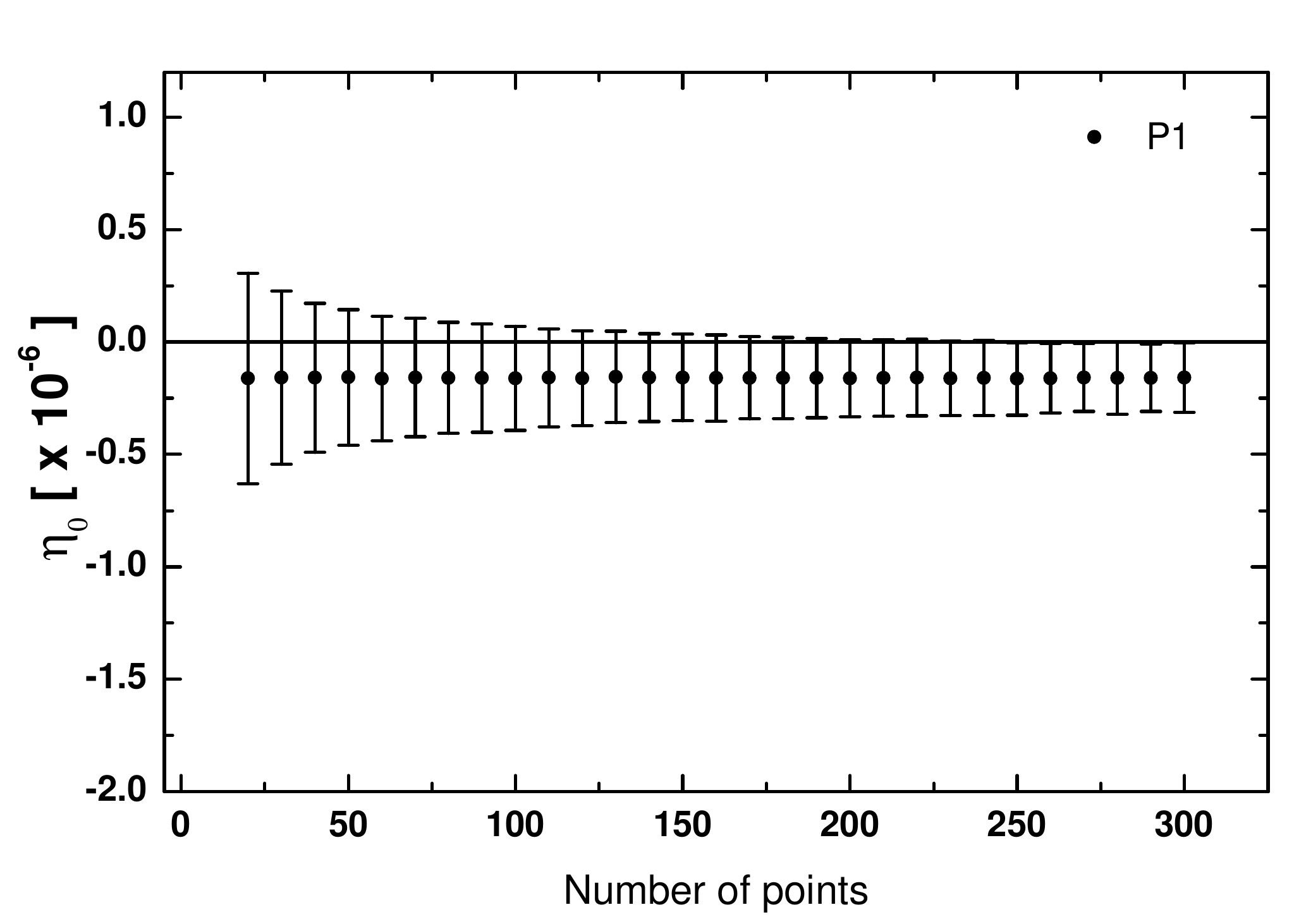}  
\includegraphics[width=0.329\textwidth]{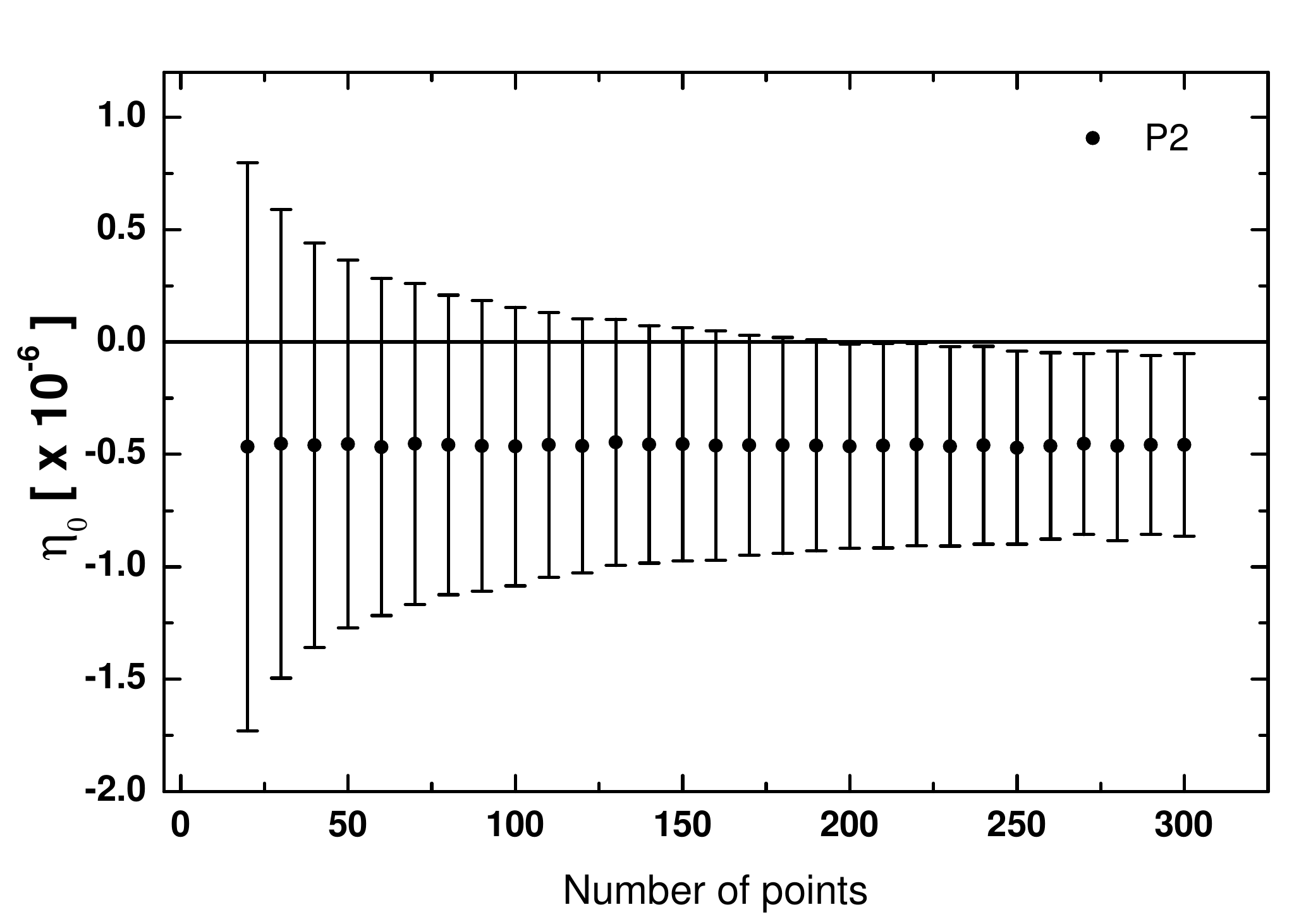} 
\includegraphics[width=0.329\textwidth]{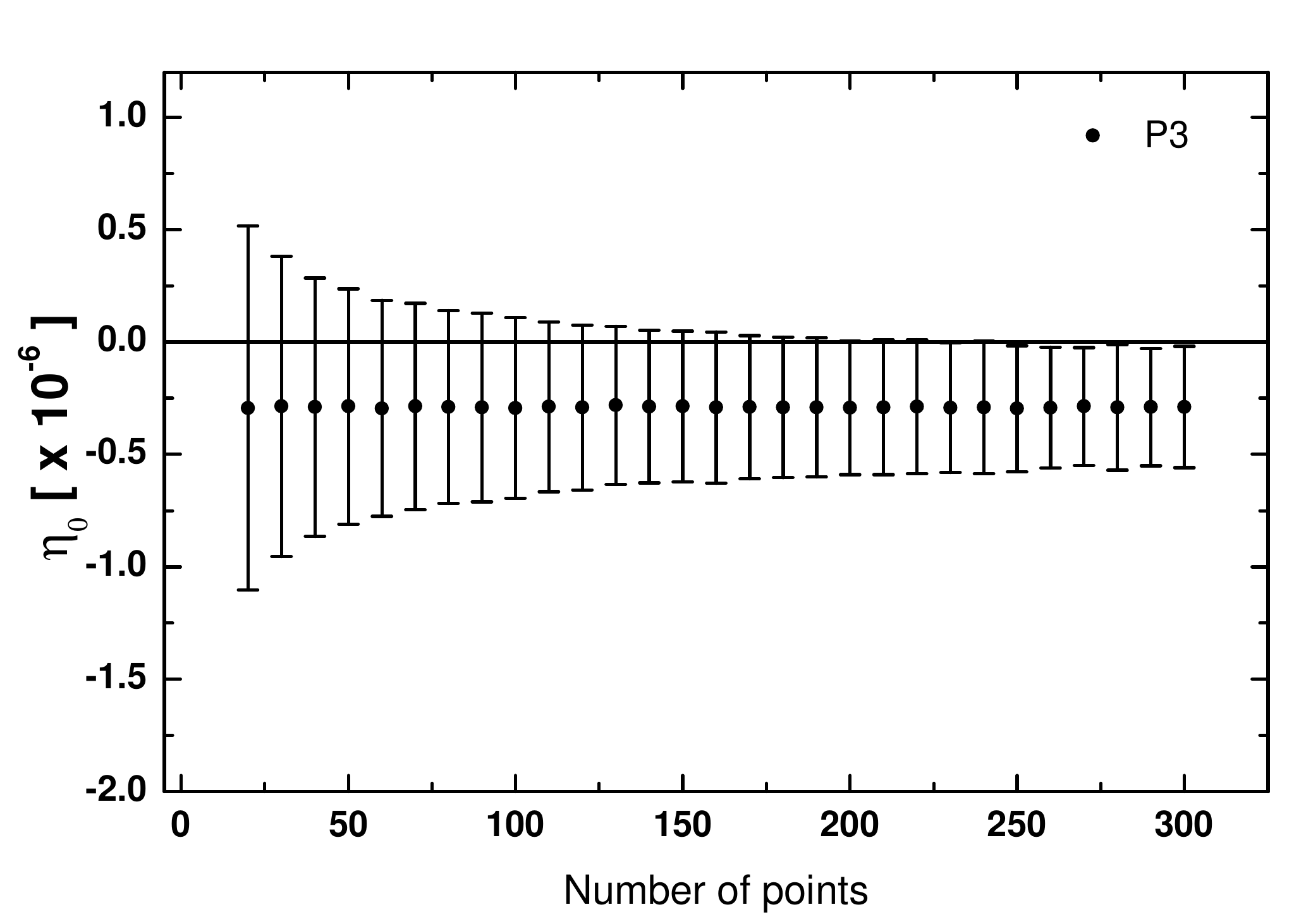}  
\caption{Simulated evolution of $\eta_0$ with the number of data points for parameterization P1.}
\label{P1evol}
\end{figure*}


\begin{table*}[!t]
\centering
\begin{tabular}{|c|c|c|c|c|}
\hline
Pn \,\,& \,\, VLT (baseline) [$\times 10^{-7}$] \,\,& \,\, VLT (ideal) [$\times 10^{-7}$]\,\, & \,\, E-ELT (baseline) [$\times 10^{-7}$] \,\,& \,\, E-ELT (ideal) [$\times 10^{-7}$]\,\,\\
\hline
P1  \,\,& \,\, $2.77 \pm 3.89 \times 10^{-1}$ \,\,& \,\, $2.77 \pm 1.34 \times 10^{-1}$ \,\,& \,\, $2.77 \pm 7.51 \times 10^{-2}$ \,\,& \,\, $2.77 \pm 4.36 \times 10^{-2}$ \\
P2  \,\,& \,\, $3.26 \pm 7.23 \times 10^{-1}$ \,\,& \,\, $3.26 \pm 2.45 \times 10^{-1}$ \,\,& \,\, $3.26 \pm 1.29 \times 10^{-1}$ \,\,& \,\, $3.26 \pm 5.85 \times 10^{-2}$ \\
P3  \,\,& \,\, $4.09 \pm 7.24 \times 10^{-1}$ \,\,& \,\, $4.09 \pm 2.47 \times 10^{-1}$ \,\,& \,\, $4.09 \pm 1.33 \times 10^{-1}$ \,\,& \,\, $4.09 \pm 6.81 \times 10^{-2}$ \\
P4  \,\,& \,\, $4.09 \pm 7.24 \times 10^{-1}$ \,\,& \,\, $4.09 \pm 2.47 \times 10^{-1}$ \,\,& \,\, $4.09 \pm 1.33 \times 10^{-1}$ \,\,& \,\, $4.09 \pm 6.81 \times 10^{-2}$ \\
\hline
\end{tabular}
\caption{Estimates of $\eta_0$ obtained from the targets and estimated uncertainties of  ESPRESSO/VLT and ELT/HIRES}. 
\label{TableIV}
\end{table*}

\begin{table*}[!t]
\centering
\begin{tabular}{|c|c|c|c|c|}
\hline
$N_{sim}$ \,\,& \,\, VLT (baseline) [$\times 10^{-7}$] \,\,& \,\, VLT (ideal) [$\times 10^{-7}$]\,\, & \,\, E-ELT (baseline) [$\times 10^{-7}$] \,\,& \,\, E-ELT (ideal) [$\times 10^{-7}$]\,\,\\
\hline
15  \,\,& \,\, $2.77 \pm 3.28 \times 10^{-1}$ \,\,& \,\, $2.77 \pm 1.15 \times 10^{-1}$ \,\,& \,\, $2.77 \pm 6.69 \times 10^{-2}$ \,\,& \,\, $2.77 \pm 4.24 \times 10^{-2}$ \\
20  \,\,& \,\, $2.77 \pm 2.85 \times 10^{-1}$ \,\,& \,\, $2.77 \pm 9.49 \times 10^{-2}$ \,\,& \,\, $2.77 \pm 4.74 \times 10^{-2}$ \,\,& \,\, $2.77 \pm 1.42 \times 10^{-2}$ \\
25  \,\,& \,\, $2.77 \pm 2.54 \times 10^{-1}$ \,\,& \,\, $2.77 \pm 8.45 \times 10^{-2}$ \,\,& \,\, $2.77 \pm 4.23 \times 10^{-2}$ \,\,& \,\, $4.09 \pm 1.27 \times 10^{-2}$ \\
30  \,\,& \,\, $2.77 \pm 2.29 \times 10^{-1}$ \,\,& \,\, $2.77 \pm 7.64 \times 10^{-2}$ \,\,& \,\, $2.77 \pm 3.82 \times 10^{-2}$ \,\,& \,\, $4.09 \pm 1.15 \times 10^{-2}$ \\
\hline
\end{tabular}
\caption{Estimates of $\eta_0$ for parameterization P1 for different values of total number of measurements. The first column shows the number of measurements. The second and third columnn show the results for the ESPRESSO/VLT baseline and ideal scenario, respectively, while the fourth and fifth columns show the results for the E-ELT-HIRES baseline and ideal scenario, respectively. }
\label{resp1}
\end{table*}

\begin{table*}[!t]
\centering
\begin{tabular}{|c|c|c|c|c|}
\hline
Pn \,\,& \,\, VLT (baseline) [$\times 10^{-7}$] \,\,& \,\, VLT (ideal) [$\times 10^{-7}$]\,\, & \,\, E-ELT (baseline) [$\times 10^{-7}$] \,\,& \,\, E-ELT (ideal) [$\times 10^{-7}$]\,\,\\
\hline
15  \,\,& \,\,  $3.13 \pm 6.53 \times 10^{-1}$ \,\,& \,\, $3.13 \pm 2.22 \times 10^{-1}$ \,\,& \,\, $3.13 \pm 1.17 \times 10^{-1}$ \,\,& \,\, $3.13 \pm 5.49 \times 10^{-2}$ \\
20  \,\,& \,\,  $3.13 \pm 5.69 \times 10^{-1}$ \,\,& \,\, $3.13 \pm 1.90 \times 10^{-1}$ \,\,& \,\, $3.13 \pm 9.49 \times 10^{-2}$ \,\,& \,\, $3.13 \pm 2.85 \times 10^{-2}$ \\ 
25  \,\,& \,\,  $3.13 \pm 5.06 \times 10^{-1}$ \,\,& \,\, $3.13 \pm 1.69 \times 10^{-1}$ \,\,& \,\, $3.13 \pm 8.43 \times 10^{-2}$ \,\,& \,\, $3.13 \pm 2.53 \times 10^{-2}$ \\ 
30  \,\,& \,\,  $3.13 \pm 4.57 \times 10^{-1}$ \,\,& \,\, $3.13 \pm 1.52 \times 10^{-1}$ \,\,& \,\, $3.13 \pm 7.62 \times 10^{-2}$ \,\,& \,\, $3.13 \pm 2.29 \times 10^{-2}$ \\
\hline
\end{tabular}
\caption{The same as in Table \ref{resp1} for P2.}
\label{resp2}
\end{table*}

\begin{table*}[!t]
\centering
\begin{tabular}{|c|c|c|c|c|}
\hline
Pn \,\,& \,\, VLT (baseline) [$\times 10^{-7}$] \,\,& \,\, VLT (ideal) [$\times 10^{-7}$]\,\, & \,\, E-ELT (baseline) [$\times 10^{-7}$] \,\,& \,\, E-ELT (ideal) [$\times 10^{-7}$]\,\,\\
\hline
15   \,\,& \,\, $4.09   \pm 6.54 \times 10^{-1}$ \,\,& \,\, $4.09   \pm 2.25 \times 10^{-1}$ \,\,& \,\,  $4.09   \pm 1.23 \times 10^{-2}$ \,\,& \,\,  $4.09 \pm 6.63 \times 10^{-2}$\\
20   \,\,& \,\, $4.09   \pm 5.90 \times 10^{-1}$ \,\,& \,\, $4.09   \pm 1.90 \times 10^{-1}$ \,\,& \,\,  $4.09   \pm 9.49 \times 10^{-2}$ \,\,& \,\,  $4.09 \pm 2.85 \times 10^{-2}$\\
25   \,\,& \,\, $4.09   \pm 5.06 \times 10^{-1}$ \,\,& \,\, $4.09   \pm 1.69 \times 10^{-1}$ \,\,& \,\,  $4.09   \pm 8.43 \times 10^{-2}$ \,\,& \,\,  $4.09 \pm 2.53 \times 10^{-2}$\\
30   \,\,& \,\, $4.09   \pm 4.57 \times 10^{-1}$ \,\,& \,\, $4.09   \pm 1.52 \times 10^{-1}$ \,\,& \,\,  $4.09   \pm 7.62 \times 10^{-2}$ \,\,& \,\,  $4.09 \pm 2.29 \times 10^{-2}$\\
\hline
\end{tabular}
\caption{The same as in Table \ref{resp1} for P3.}
\label{resp3}	
\end{table*}

\section{Forecast Analysis}
\label{stats}

In this section we describe the forecast analysis performed with the inferred observational  values of $\eta_0$ from both the currently available data set  and the targets available for future missions. We use the three parameterizations described in the previous section and show that the results are independent of the parameterization adopted. 

We performed a series of Monte Carlo simulations, divided into two sets of forecast (Forecast I and Forecast II), which are explained below. In order to make clearer the simulation procedure we describe the general methodology as follows:

\begin{itemize}

\item We set the fiducial model of $\eta(z)$ for each parameterization with the corresponding best-fit value of $\eta_0$ obtained from current observations (Table \ref{TableIII});

\item {For Forecast I}, we use the relative error ($\sigma_{\eta(z)}/\eta(z)$) obtained from the observational data. For Forecast II, we consider the mean value of $\eta$ from the observational inferred values shown in Table \ref{DS2} and the uncertainties as described in Section \ref{obs} for future surveys. We then perform a statistical analysis and obtain a linear fit from the relative error, resulting in a mean  and 1$\sigma$ errors. We remove the points that are further than 1$\sigma$ from the previous fit and, with the remaining points, we perform again the same statistical analysis to obtain the final mean and 1$\sigma$ errors. 

\item By the convolution of the final fit for the relative error with the fiducial model we obtain the final fit for the error with a mean and 1$\sigma$ errors;

\item With the fiducial model and errors, we run the simulations by setting the number of points in each realisation of the simulations. The points are equally distributed in redshift;

\item The mean value of the simulation, for each redshift, is obtained from a random pick in a Gaussian distribution ($\cal N$ ($\mu$,$\sigma$)) with the mean ($\mu$) and standard deviation ($\sigma$), given by the fiducial model;

\item The corresponding errors of the simulation, for each redshift, is also obtained from a random pick in the Gaussian distribution $\cal N$ ($\mu$,$\sigma$). In this case, with the mean ($\mu$) and standard deviation ($\sigma$), as previously described, given by the convolution of the final fit for the relative error with the fiducial model;

\item The previous procedures are repeated for each redshift. 

\item A $\chi^2$ statistics is used to obtain the mean best fit value of $\eta_0$ for a given realisation.

\item The previous procedure is repeated for a number of realisations ($\sim O(10^4)$), providing an average and standard deviation of the best fit values. Such values are taken as the mean value and error of $\eta_0$, given the number of points in the simulated set.

\item The whole procedure is repeated for different numbers of points.
	
\end{itemize}


\subsection{Forecast I} 

For exemplification, we show in Fig.~\ref{Chi2Pars} (right) a comparison between the observed data points of Table \ref{DS1} (black points) and simulated points of a given realisation from the forecast I (blue points), assuming P1. The evolution of the values of $\eta_0$ with the number of points $N$ assumed in the simulations is shown in Figs. \ref{P1evol} for  parameterizations P1 (left), P2 (middle) and  P3 (right). As expected, the larger the number of points, the smaller the errors obtained. Thus, we define the critical number, $N_{crit}$, at which $\sigma_{\eta_0} = \eta_0$ and the validity of the DDR is excluded within 1$\sigma$. For parameterizations P1, P2, and P3 we find $N_{crit} = 248$, $N_{crit} = 195$, $N_{crit} = 226$, respectively. It should be noted that the value of  $N_{crit}$ does not depend significantly on the parameterization adopted.

\subsection{Forecast II} 

The previous forecast used the observationally inferred data set for $\eta(z)$ obtained from the currently available sample of ${\Delta \alpha}/{\alpha}$  (Table \ref{DS1}) as the starting point of the simulations. In our second forecast analysis, we investigate the ability of future observations to constrain the DDR parameter $\eta(z)$. We use the redshift range of Table \ref{DS2} and obtain the mean values of $\eta$ applying Eq. (\ref{FromAlphaToEta}) to Table \ref{DS2}. We also take the errors equal to  the uncertainty expectations of the missions ESPRESSO/VLT and ELT/HIRES for the baseline and ideal scenario, as described in Section \ref{obs}.

We first perform the data simulation and forecast analyses by fixing the number of points to the DS2 data set. The results  for each parameterization (with $N = N_{DS2}$) are presented in Table \ref{TableIV}. Then, we perform the same analysis by varying the number of measurements. The results for parametrizations P1, P2 and P3 are shown respectively in Tables \ref{resp1},\ref{resp2} and \ref{resp3}. Our analysis shows that with the uncertainty level of the next generation of high-resolution spectrograph (ESPRESSO/VLT and E-ELT/HIRES) the validity of the DRR can be verified with only $15$ measurements, i.e., one order of magnitude lower than the number of data points required by the current data. Note also that the expected bounds on $\eta_0$ ($O (10^{-9})$) are two orders of magnitude lower than the current ones.  Such conclusion holds for all parameterizations adopted in this paper.

\section{Conclusions}
\label{conclusions}

The search for space-time dependence of the fundamental constants of Nature is crucial to investigate potential signs of new physics. In this paper we used the best currently available measurements of ${\Delta \alpha}/{\alpha}$ to impose the most stringent constraints on a possible violation of the DDR to date ($\eta_0 \simeq O(10^{-7})$), assuming a direct relation between variations of $\alpha$ and departures of the DDR ($\eta = 1$), as derived in \cite{Hees2014,Minazzoli2014}.

Furthermore, we used the estimates of uncertainty of upcoming missions, such as VLT/ESPRESSO and E-ELT/HIRES, to forecast future constraints on the DDR parameter and estimate the necessary number of data points to confirm the hypotheses behind the DDR. Our results show that, for the level of uncertainties of the present data set, a number of observations $N_{crit} \gtrsim 200$ is enough to verify the validity of the DDR whereas for the expected level of uncertainties of the upcoming measurements, we found  $N_{crit} \sim 15$. The expected bounds on $\eta_0$ are of the order of $10^{-9}$, regardless of the DDR parameterization adopted.

\section*{Acknowledgments}

RSG is supported by CAPES/Brazil through a PNPD fellowship. S.L. is supported  by the National Agency for the Promotion of Science and Technology (ANPCYT) of Argentina grant PICT-2016-0081; by grant G140 from UNLP and by grant UBACYT 20020170100129BA. JSA acknowledges support from FAPERJ (Rio de Janeiro Science Foundation) grant no. E-26/203.024/2017 and CNPq/Brazil grant No. 310790/2014-0 and No. 400471/2014-0. RFLH acknowledges financial support from CNPq/Brazil (No. 303734/2014-0).

\bibliography{bibliov2}

\begin{thebibliography}{69}%
\makeatletter
\providecommand \@ifxundefined [1]{%
 \@ifx{#1\undefined}
}%
\providecommand \@ifnum [1]{%
 \ifnum #1\expandafter \@firstoftwo
 \else \expandafter \@secondoftwo
 \fi
}%
\providecommand \@ifx [1]{%
 \ifx #1\expandafter \@firstoftwo
 \else \expandafter \@secondoftwo
 \fi
}%
\providecommand \natexlab [1]{#1}%
\providecommand \enquote  [1]{``#1''}%
\providecommand \bibnamefont  [1]{#1}%
\providecommand \bibfnamefont [1]{#1}%
\providecommand \citenamefont [1]{#1}%
\providecommand \href@noop [0]{\@secondoftwo}%
\providecommand \href [0]{\begingroup \@sanitize@url \@href}%
\providecommand \@href[1]{\@@startlink{#1}\@@href}%
\providecommand \@@href[1]{\endgroup#1\@@endlink}%
\providecommand \@sanitize@url [0]{\catcode `\\12\catcode `\$12\catcode
  `\&12\catcode `\#12\catcode `\^12\catcode `\_12\catcode `\%12\relax}%
\providecommand \@@startlink[1]{}%
\providecommand \@@endlink[0]{}%
\providecommand \url  [0]{\begingroup\@sanitize@url \@url }%
\providecommand \@url [1]{\endgroup\@href {#1}{\urlprefix }}%
\providecommand \urlprefix  [0]{URL }%
\providecommand \Eprint [0]{\href }%
\providecommand \doibase [0]{http://dx.doi.org/}%
\providecommand \selectlanguage [0]{\@gobble}%
\providecommand \bibinfo  [0]{\@secondoftwo}%
\providecommand \bibfield  [0]{\@secondoftwo}%
\providecommand \translation [1]{[#1]}%
\providecommand \BibitemOpen [0]{}%
\providecommand \bibitemStop [0]{}%
\providecommand \bibitemNoStop [0]{.\EOS\space}%
\providecommand \EOS [0]{\spacefactor3000\relax}%
\providecommand \BibitemShut  [1]{\csname bibitem#1\endcsname}%
\let\auto@bib@innerbib\@empty
\bibitem [{\citenamefont {{Dirac}}(1937)}]{dirac1}%
  \BibitemOpen
  \bibfield  {author} {\bibinfo {author} {\bibfnamefont {P.~A.~M.}\
  \bibnamefont {{Dirac}}},\ }\href {\doibase 10.1038/139323a0} {\bibfield
  {journal} {\bibinfo  {journal} {\nat}\ }\textbf {\bibinfo {volume} {139}},\
  \bibinfo {pages} {323} (\bibinfo {year} {1937})}\BibitemShut {NoStop}%
\bibitem [{\citenamefont {{Uzan}}(2011)}]{Uzan}%
  \BibitemOpen
  \bibfield  {author} {\bibinfo {author} {\bibfnamefont {J.-P.}\ \bibnamefont
  {{Uzan}}},\ }\href {\doibase 10.12942/lrr-2011-2} {\bibfield  {journal}
  {\bibinfo  {journal} {Living Reviews in Relativity}\ }\textbf {\bibinfo
  {volume} {14}},\ \bibinfo {pages} {2} (\bibinfo {year} {2011})},\ \Eprint
  {http://arxiv.org/abs/1009.5514} {arXiv:1009.5514 [astro-ph.CO]} \BibitemShut
  {NoStop}%
\bibitem [{\citenamefont {{Martins}}(2017)}]{Martins2017}%
  \BibitemOpen
  \bibfield  {author} {\bibinfo {author} {\bibfnamefont {C.~J.~A.~P.}\
  \bibnamefont {{Martins}}},\ }\href {\doibase 10.1088/1361-6633/aa860e}
  {\bibfield  {journal} {\bibinfo  {journal} {Reports on Progress in Physics}\
  }\textbf {\bibinfo {volume} {80}},\ \bibinfo {eid} {126902} (\bibinfo {year}
  {2017})},\ \Eprint {http://arxiv.org/abs/1709.02923} {arXiv:1709.02923
  [astro-ph.CO]} \BibitemShut {NoStop}%
\bibitem [{\citenamefont {{Damour}}\ and\ \citenamefont
  {{Dyson}}(1996)}]{1996NuPhB.480...37D}%
  \BibitemOpen
  \bibfield  {author} {\bibinfo {author} {\bibfnamefont {T.}~\bibnamefont
  {{Damour}}}\ and\ \bibinfo {author} {\bibfnamefont {F.}~\bibnamefont
  {{Dyson}}},\ }\href {\doibase 10.1016/S0550-3213(96)00467-1} {\bibfield
  {journal} {\bibinfo  {journal} {Nuclear Physics B}\ }\textbf {\bibinfo
  {volume} {480}},\ \bibinfo {pages} {37} (\bibinfo {year} {1996})},\ \Eprint
  {http://arxiv.org/abs/hep-ph/9606486} {arXiv:hep-ph/9606486 [hep-ph]}
  \BibitemShut {NoStop}%
\bibitem [{\citenamefont {{Petrov}}\ \emph {et~al.}(2006)\citenamefont
  {{Petrov}}, \citenamefont {{Nazarov}}, \citenamefont {{Onegin}},
  \citenamefont {{Petrov}},\ and\ \citenamefont {{Sakhnovsky}}}]{Petrov06}%
  \BibitemOpen
  \bibfield  {author} {\bibinfo {author} {\bibfnamefont {Y.~V.}\ \bibnamefont
  {{Petrov}}}, \bibinfo {author} {\bibfnamefont {A.~I.}\ \bibnamefont
  {{Nazarov}}}, \bibinfo {author} {\bibfnamefont {M.~S.}\ \bibnamefont
  {{Onegin}}}, \bibinfo {author} {\bibfnamefont {V.~Y.}\ \bibnamefont
  {{Petrov}}}, \ and\ \bibinfo {author} {\bibfnamefont {E.~G.}\ \bibnamefont
  {{Sakhnovsky}}},\ }\href {\doibase 10.1103/PhysRevC.74.064610} {\bibfield
  {journal} {\bibinfo  {journal} {\prc}\ }\textbf {\bibinfo {volume} {74}},\
  \bibinfo {eid} {064610} (\bibinfo {year} {2006})},\ \Eprint
  {http://arxiv.org/abs/hep-ph/0506186} {hep-ph/0506186} \BibitemShut {NoStop}%
\bibitem [{\citenamefont {{Gould}}\ \emph {et~al.}(2006)\citenamefont
  {{Gould}}, \citenamefont {{Sharapov}},\ and\ \citenamefont
  {{Lamoreaux}}}]{Gould06}%
  \BibitemOpen
  \bibfield  {author} {\bibinfo {author} {\bibfnamefont {C.~R.}\ \bibnamefont
  {{Gould}}}, \bibinfo {author} {\bibfnamefont {E.~I.}\ \bibnamefont
  {{Sharapov}}}, \ and\ \bibinfo {author} {\bibfnamefont {S.~K.}\ \bibnamefont
  {{Lamoreaux}}},\ }\href {\doibase 10.1103/PhysRevC.74.024607} {\bibfield
  {journal} {\bibinfo  {journal} {\prc}\ }\textbf {\bibinfo {volume} {74}},\
  \bibinfo {eid} {024607} (\bibinfo {year} {2006})},\ \Eprint
  {http://arxiv.org/abs/nucl-ex/0701019} {nucl-ex/0701019} \BibitemShut
  {NoStop}%
\bibitem [{\citenamefont {Dyson}(1967)}]{dyson}%
  \BibitemOpen
  \bibfield  {author} {\bibinfo {author} {\bibfnamefont {F.~J.}\ \bibnamefont
  {Dyson}},\ }\href {\doibase 10.1103/PhysRevLett.19.1291} {\bibfield
  {journal} {\bibinfo  {journal} {Phys. Rev. Lett.}\ }\textbf {\bibinfo
  {volume} {19}},\ \bibinfo {pages} {1291} (\bibinfo {year}
  {1967})}\BibitemShut {NoStop}%
\bibitem [{\citenamefont {Sisterna}\ and\ \citenamefont
  {Vucetich}(1990)}]{sisterna}%
  \BibitemOpen
  \bibfield  {author} {\bibinfo {author} {\bibfnamefont {P.}~\bibnamefont
  {Sisterna}}\ and\ \bibinfo {author} {\bibfnamefont {H.}~\bibnamefont
  {Vucetich}},\ }\href {\doibase 10.1103/PhysRevD.41.1034} {\bibfield
  {journal} {\bibinfo  {journal} {Phys. Rev. D}\ }\textbf {\bibinfo {volume}
  {41}},\ \bibinfo {pages} {1034} (\bibinfo {year} {1990})}\BibitemShut
  {NoStop}%
\bibitem [{\citenamefont {{Olive}}\ \emph {et~al.}(2004)\citenamefont
  {{Olive}}, \citenamefont {{Pospelov}}, \citenamefont {{Qian}}, \citenamefont
  {{Manh{\`e}s}}, \citenamefont {{Vangioni-Flam}}, \citenamefont {{Coc}},\ and\
  \citenamefont {{Cass{\'e}}}}]{Olive04b}%
  \BibitemOpen
  \bibfield  {author} {\bibinfo {author} {\bibfnamefont {K.~A.}\ \bibnamefont
  {{Olive}}}, \bibinfo {author} {\bibfnamefont {M.}~\bibnamefont {{Pospelov}}},
  \bibinfo {author} {\bibfnamefont {Y.-Z.}\ \bibnamefont {{Qian}}}, \bibinfo
  {author} {\bibfnamefont {G.}~\bibnamefont {{Manh{\`e}s}}}, \bibinfo {author}
  {\bibfnamefont {E.}~\bibnamefont {{Vangioni-Flam}}}, \bibinfo {author}
  {\bibfnamefont {A.}~\bibnamefont {{Coc}}}, \ and\ \bibinfo {author}
  {\bibfnamefont {M.}~\bibnamefont {{Cass{\'e}}}},\ }\href {\doibase
  10.1103/PhysRevD.69.027701} {\bibfield  {journal} {\bibinfo  {journal}
  {\prd}\ }\textbf {\bibinfo {volume} {69}},\ \bibinfo {eid} {027701} (\bibinfo
  {year} {2004})},\ \Eprint {http://arxiv.org/abs/astro-ph/0309252}
  {astro-ph/0309252} \BibitemShut {NoStop}%
\bibitem [{\citenamefont {{Prestage}}\ \emph {et~al.}(1995)\citenamefont
  {{Prestage}}, \citenamefont {{Tjoelker}},\ and\ \citenamefont
  {{Maleki}}}]{pres}%
  \BibitemOpen
  \bibfield  {author} {\bibinfo {author} {\bibfnamefont {J.~D.}\ \bibnamefont
  {{Prestage}}}, \bibinfo {author} {\bibfnamefont {R.~L.}\ \bibnamefont
  {{Tjoelker}}}, \ and\ \bibinfo {author} {\bibfnamefont {L.}~\bibnamefont
  {{Maleki}}},\ }\href {\doibase 10.1103/PhysRevLett.74.3511} {\bibfield
  {journal} {\bibinfo  {journal} {Physical Review Letters}\ }\textbf {\bibinfo
  {volume} {74}},\ \bibinfo {pages} {3511} (\bibinfo {year}
  {1995})}\BibitemShut {NoStop}%
\bibitem [{\citenamefont {{Peik}}\ \emph {et~al.}(2004)\citenamefont {{Peik}},
  \citenamefont {{Lipphardt}}, \citenamefont {{Schnatz}}, \citenamefont
  {{Schneider}}, \citenamefont {{Tamm}},\ and\ \citenamefont
  {{Karshenboim}}}]{Peik04}%
  \BibitemOpen
  \bibfield  {author} {\bibinfo {author} {\bibfnamefont {E.}~\bibnamefont
  {{Peik}}}, \bibinfo {author} {\bibfnamefont {B.}~\bibnamefont {{Lipphardt}}},
  \bibinfo {author} {\bibfnamefont {H.}~\bibnamefont {{Schnatz}}}, \bibinfo
  {author} {\bibfnamefont {T.}~\bibnamefont {{Schneider}}}, \bibinfo {author}
  {\bibfnamefont {C.}~\bibnamefont {{Tamm}}}, \ and\ \bibinfo {author}
  {\bibfnamefont {S.~G.}\ \bibnamefont {{Karshenboim}}},\ }\href {\doibase
  10.1103/PhysRevLett.93.170801} {\bibfield  {journal} {\bibinfo  {journal}
  {Physical Review Letters}\ }\textbf {\bibinfo {volume} {93}},\ \bibinfo {eid}
  {170801} (\bibinfo {year} {2004})},\ \Eprint
  {http://arxiv.org/abs/physics/0402132} {physics/0402132} \BibitemShut
  {NoStop}%
\bibitem [{\citenamefont {{Rosenband}}\ \emph {et~al.}(2008)\citenamefont
  {{Rosenband}}, \citenamefont {{Hume}}, \citenamefont {{Schmidt}},
  \citenamefont {{Chou}}, \citenamefont {{Brusch}}, \citenamefont {{Lorini}},
  \citenamefont {{Oskay}}, \citenamefont {{Drullinger}}, \citenamefont
  {{Fortier}}, \citenamefont {{Stalnaker}}, \citenamefont {{Diddams}},
  \citenamefont {{Swann}}, \citenamefont {{Newbury}}, \citenamefont {{Itano}},
  \citenamefont {{Wineland}},\ and\ \citenamefont {{Bergquist}}}]{rosenband08}%
  \BibitemOpen
  \bibfield  {author} {\bibinfo {author} {\bibfnamefont {T.}~\bibnamefont
  {{Rosenband}}}, \bibinfo {author} {\bibfnamefont {D.~B.}\ \bibnamefont
  {{Hume}}}, \bibinfo {author} {\bibfnamefont {P.~O.}\ \bibnamefont
  {{Schmidt}}}, \bibinfo {author} {\bibfnamefont {C.~W.}\ \bibnamefont
  {{Chou}}}, \bibinfo {author} {\bibfnamefont {A.}~\bibnamefont {{Brusch}}},
  \bibinfo {author} {\bibfnamefont {L.}~\bibnamefont {{Lorini}}}, \bibinfo
  {author} {\bibfnamefont {W.~H.}\ \bibnamefont {{Oskay}}}, \bibinfo {author}
  {\bibfnamefont {R.~E.}\ \bibnamefont {{Drullinger}}}, \bibinfo {author}
  {\bibfnamefont {T.~M.}\ \bibnamefont {{Fortier}}}, \bibinfo {author}
  {\bibfnamefont {J.~E.}\ \bibnamefont {{Stalnaker}}}, \bibinfo {author}
  {\bibfnamefont {S.~A.}\ \bibnamefont {{Diddams}}}, \bibinfo {author}
  {\bibfnamefont {W.~C.}\ \bibnamefont {{Swann}}}, \bibinfo {author}
  {\bibfnamefont {N.~R.}\ \bibnamefont {{Newbury}}}, \bibinfo {author}
  {\bibfnamefont {W.~M.}\ \bibnamefont {{Itano}}}, \bibinfo {author}
  {\bibfnamefont {D.~J.}\ \bibnamefont {{Wineland}}}, \ and\ \bibinfo {author}
  {\bibfnamefont {J.~C.}\ \bibnamefont {{Bergquist}}},\ }\href {\doibase
  10.1126/science.1154622} {\bibfield  {journal} {\bibinfo  {journal}
  {Science}\ }\textbf {\bibinfo {volume} {319}},\ \bibinfo {pages} {1808}
  (\bibinfo {year} {2008})}\BibitemShut {NoStop}%
\bibitem [{\citenamefont {{Bahcall}}\ \emph {et~al.}(2004)\citenamefont
  {{Bahcall}}, \citenamefont {{Steinhardt}},\ and\ \citenamefont
  {{Schlegel}}}]{bahc}%
  \BibitemOpen
  \bibfield  {author} {\bibinfo {author} {\bibfnamefont {J.~N.}\ \bibnamefont
  {{Bahcall}}}, \bibinfo {author} {\bibfnamefont {C.~L.}\ \bibnamefont
  {{Steinhardt}}}, \ and\ \bibinfo {author} {\bibfnamefont {D.}~\bibnamefont
  {{Schlegel}}},\ }\href {\doibase 10.1086/379971} {\bibfield  {journal}
  {\bibinfo  {journal} {\apj}\ }\textbf {\bibinfo {volume} {600}},\ \bibinfo
  {pages} {520} (\bibinfo {year} {2004})},\ \Eprint
  {http://arxiv.org/abs/astro-ph/0301507} {astro-ph/0301507} \BibitemShut
  {NoStop}%
\bibitem [{\citenamefont {{Levshakov}}\ \emph {et~al.}(2002)\citenamefont
  {{Levshakov}}, \citenamefont {{Dessauges-Zavadsky}}, \citenamefont
  {{D'Odorico}},\ and\ \citenamefont {{Molaro}}}]{lev}%
  \BibitemOpen
  \bibfield  {author} {\bibinfo {author} {\bibfnamefont {S.~A.}\ \bibnamefont
  {{Levshakov}}}, \bibinfo {author} {\bibfnamefont {M.}~\bibnamefont
  {{Dessauges-Zavadsky}}}, \bibinfo {author} {\bibfnamefont {S.}~\bibnamefont
  {{D'Odorico}}}, \ and\ \bibinfo {author} {\bibfnamefont {P.}~\bibnamefont
  {{Molaro}}},\ }\href {\doibase 10.1046/j.1365-8711.2002.05408.x} {\bibfield
  {journal} {\bibinfo  {journal} {Monthly Notices Royal Astronomical Society}\
  }\textbf {\bibinfo {volume} {333}},\ \bibinfo {pages} {373} (\bibinfo {year}
  {2002})},\ \Eprint {http://arxiv.org/abs/astro-ph/0106194} {astro-ph/0106194}
  \BibitemShut {NoStop}%
\bibitem [{\citenamefont {{Murphy}}\ \emph
  {et~al.}(2001{\natexlab{a}})\citenamefont {{Murphy}}, \citenamefont {{Webb}},
  \citenamefont {{Flambaum}}, \citenamefont {{Dzuba}}, \citenamefont
  {{Churchill}}, \citenamefont {{Prochaska}}, \citenamefont {{Barrow}},\ and\
  \citenamefont {{Wolfe}}}]{murphy1}%
  \BibitemOpen
  \bibfield  {author} {\bibinfo {author} {\bibfnamefont {M.~T.}\ \bibnamefont
  {{Murphy}}}, \bibinfo {author} {\bibfnamefont {J.~K.}\ \bibnamefont
  {{Webb}}}, \bibinfo {author} {\bibfnamefont {V.~V.}\ \bibnamefont
  {{Flambaum}}}, \bibinfo {author} {\bibfnamefont {V.~A.}\ \bibnamefont
  {{Dzuba}}}, \bibinfo {author} {\bibfnamefont {C.~W.}\ \bibnamefont
  {{Churchill}}}, \bibinfo {author} {\bibfnamefont {J.~X.}\ \bibnamefont
  {{Prochaska}}}, \bibinfo {author} {\bibfnamefont {J.~D.}\ \bibnamefont
  {{Barrow}}}, \ and\ \bibinfo {author} {\bibfnamefont {A.~M.}\ \bibnamefont
  {{Wolfe}}},\ }\href {\doibase 10.1046/j.1365-8711.2001.04840.x} {\bibfield
  {journal} {\bibinfo  {journal} {Monthly Notices Royal Astronomical Society}\
  }\textbf {\bibinfo {volume} {327}},\ \bibinfo {pages} {1208} (\bibinfo {year}
  {2001}{\natexlab{a}})},\ \Eprint {http://arxiv.org/abs/astro-ph/0012419}
  {astro-ph/0012419} \BibitemShut {NoStop}%
\bibitem [{\citenamefont {{Murphy}}\ \emph
  {et~al.}(2001{\natexlab{b}})\citenamefont {{Murphy}}, \citenamefont {{Webb}},
  \citenamefont {{Flambaum}}, \citenamefont {{Prochaska}},\ and\ \citenamefont
  {{Wolfe}}}]{murphy2}%
  \BibitemOpen
  \bibfield  {author} {\bibinfo {author} {\bibfnamefont {M.~T.}\ \bibnamefont
  {{Murphy}}}, \bibinfo {author} {\bibfnamefont {J.~K.}\ \bibnamefont
  {{Webb}}}, \bibinfo {author} {\bibfnamefont {V.~V.}\ \bibnamefont
  {{Flambaum}}}, \bibinfo {author} {\bibfnamefont {J.~X.}\ \bibnamefont
  {{Prochaska}}}, \ and\ \bibinfo {author} {\bibfnamefont {A.~M.}\ \bibnamefont
  {{Wolfe}}},\ }\href {\doibase 10.1046/j.1365-8711.2001.04842.x} {\bibfield
  {journal} {\bibinfo  {journal} {Monthly Notices Royal Astronomical Society}\
  }\textbf {\bibinfo {volume} {327}},\ \bibinfo {pages} {1237} (\bibinfo {year}
  {2001}{\natexlab{b}})},\ \Eprint {http://arxiv.org/abs/astro-ph/0012421}
  {astro-ph/0012421} \BibitemShut {NoStop}%
\bibitem [{\citenamefont {{Webb}}\ \emph {et~al.}(1999)\citenamefont {{Webb}},
  \citenamefont {{Flambaum}}, \citenamefont {{Churchill}}, \citenamefont
  {{Drinkwater}},\ and\ \citenamefont {{Barrow}}}]{Webb99}%
  \BibitemOpen
  \bibfield  {author} {\bibinfo {author} {\bibfnamefont {J.~K.}\ \bibnamefont
  {{Webb}}}, \bibinfo {author} {\bibfnamefont {V.~V.}\ \bibnamefont
  {{Flambaum}}}, \bibinfo {author} {\bibfnamefont {C.~W.}\ \bibnamefont
  {{Churchill}}}, \bibinfo {author} {\bibfnamefont {M.~J.}\ \bibnamefont
  {{Drinkwater}}}, \ and\ \bibinfo {author} {\bibfnamefont {J.~D.}\
  \bibnamefont {{Barrow}}},\ }\href {\doibase 10.1103/PhysRevLett.82.884}
  {\bibfield  {journal} {\bibinfo  {journal} {Physical Review Letters}\
  }\textbf {\bibinfo {volume} {82}},\ \bibinfo {pages} {884} (\bibinfo {year}
  {1999})},\ \Eprint {http://arxiv.org/abs/astro-ph/9803165} {astro-ph/9803165}
  \BibitemShut {NoStop}%
\bibitem [{\citenamefont {{Webb}}\ \emph {et~al.}(2001)\citenamefont {{Webb}},
  \citenamefont {{Murphy}}, \citenamefont {{Flambaum}}, \citenamefont
  {{Dzuba}}, \citenamefont {{Barrow}}, \citenamefont {{Churchill}},
  \citenamefont {{Prochaska}},\ and\ \citenamefont {{Wolfe}}}]{webb2}%
  \BibitemOpen
  \bibfield  {author} {\bibinfo {author} {\bibfnamefont {J.~K.}\ \bibnamefont
  {{Webb}}}, \bibinfo {author} {\bibfnamefont {M.~T.}\ \bibnamefont
  {{Murphy}}}, \bibinfo {author} {\bibfnamefont {V.~V.}\ \bibnamefont
  {{Flambaum}}}, \bibinfo {author} {\bibfnamefont {V.~A.}\ \bibnamefont
  {{Dzuba}}}, \bibinfo {author} {\bibfnamefont {J.~D.}\ \bibnamefont
  {{Barrow}}}, \bibinfo {author} {\bibfnamefont {C.~W.}\ \bibnamefont
  {{Churchill}}}, \bibinfo {author} {\bibfnamefont {J.~X.}\ \bibnamefont
  {{Prochaska}}}, \ and\ \bibinfo {author} {\bibfnamefont {A.~M.}\ \bibnamefont
  {{Wolfe}}},\ }\href {\doibase 10.1103/PhysRevLett.87.091301} {\bibfield
  {journal} {\bibinfo  {journal} {Physical Review Letters}\ }\textbf {\bibinfo
  {volume} {87}},\ \bibinfo {eid} {091301} (\bibinfo {year} {2001})},\ \Eprint
  {http://arxiv.org/abs/astro-ph/0012539} {astro-ph/0012539} \BibitemShut
  {NoStop}%
\bibitem [{\citenamefont {{Galli}}(2013)}]{Galli13}%
  \BibitemOpen
  \bibfield  {author} {\bibinfo {author} {\bibfnamefont {S.}~\bibnamefont
  {{Galli}}},\ }\href {\doibase 10.1103/PhysRevD.87.123516} {\bibfield
  {journal} {\bibinfo  {journal} {\prd}\ }\textbf {\bibinfo {volume} {87}},\
  \bibinfo {eid} {123516} (\bibinfo {year} {2013})},\ \Eprint
  {http://arxiv.org/abs/1212.1075} {arXiv:1212.1075} \BibitemShut {NoStop}%
\bibitem [{\citenamefont {{Holanda}}\ \emph
  {et~al.}(2016{\natexlab{a}})\citenamefont {{Holanda}}, \citenamefont
  {{Landau}}, \citenamefont {{Alcaniz}}, \citenamefont {{S{\'a}nchez G.}},\
  and\ \citenamefont {{Busti}}}]{Holanda16}%
  \BibitemOpen
  \bibfield  {author} {\bibinfo {author} {\bibfnamefont {R.~F.~L.}\
  \bibnamefont {{Holanda}}}, \bibinfo {author} {\bibfnamefont {S.~J.}\
  \bibnamefont {{Landau}}}, \bibinfo {author} {\bibfnamefont {J.~S.}\
  \bibnamefont {{Alcaniz}}}, \bibinfo {author} {\bibfnamefont {I.~E.}\
  \bibnamefont {{S{\'a}nchez G.}}}, \ and\ \bibinfo {author} {\bibfnamefont
  {V.~C.}\ \bibnamefont {{Busti}}},\ }\href {\doibase
  10.1088/1475-7516/2016/05/047} {\bibfield  {journal} {\bibinfo  {journal}
  {Journal of Cosmology and Astroparticle Physics}\ }\textbf {\bibinfo {volume}
  {5}},\ \bibinfo {eid} {047} (\bibinfo {year} {2016}{\natexlab{a}})},\ \Eprint
  {http://arxiv.org/abs/1510.07240} {arXiv:1510.07240} \BibitemShut {NoStop}%
\bibitem [{\citenamefont {{Holanda}}\ \emph
  {et~al.}(2016{\natexlab{b}})\citenamefont {{Holanda}}, \citenamefont
  {{Busti}}, \citenamefont {{Cola{\c c}o}}, \citenamefont {{Alcaniz}},\ and\
  \citenamefont {{Landau}}}]{Holanda16b}%
  \BibitemOpen
  \bibfield  {author} {\bibinfo {author} {\bibfnamefont {R.~F.~L.}\
  \bibnamefont {{Holanda}}}, \bibinfo {author} {\bibfnamefont {V.~C.}\
  \bibnamefont {{Busti}}}, \bibinfo {author} {\bibfnamefont {L.~R.}\
  \bibnamefont {{Cola{\c c}o}}}, \bibinfo {author} {\bibfnamefont {J.~S.}\
  \bibnamefont {{Alcaniz}}}, \ and\ \bibinfo {author} {\bibfnamefont {S.~J.}\
  \bibnamefont {{Landau}}},\ }\href {\doibase 10.1088/1475-7516/2016/08/055}
  {\bibfield  {journal} {\bibinfo  {journal} {Journal of Cosmology and
  Astroparticle Physics}\ }\textbf {\bibinfo {volume} {8}},\ \bibinfo {eid}
  {055} (\bibinfo {year} {2016}{\natexlab{b}})},\ \Eprint
  {http://arxiv.org/abs/1605.02578} {arXiv:1605.02578} \BibitemShut {NoStop}%
\bibitem [{\citenamefont {{de Martino}}\ \emph
  {et~al.}(2016{\natexlab{a}})\citenamefont {{de Martino}}, \citenamefont
  {{Martins}}, \citenamefont {{Ebeling}},\ and\ \citenamefont
  {{Kocevski}}}]{Martino16}%
  \BibitemOpen
  \bibfield  {author} {\bibinfo {author} {\bibfnamefont {I.}~\bibnamefont {{de
  Martino}}}, \bibinfo {author} {\bibfnamefont {C.~J.~A.~P.}\ \bibnamefont
  {{Martins}}}, \bibinfo {author} {\bibfnamefont {H.}~\bibnamefont
  {{Ebeling}}}, \ and\ \bibinfo {author} {\bibfnamefont {D.}~\bibnamefont
  {{Kocevski}}},\ }\href {\doibase 10.1103/PhysRevD.94.083008} {\bibfield
  {journal} {\bibinfo  {journal} {\prd}\ }\textbf {\bibinfo {volume} {94}},\
  \bibinfo {eid} {083008} (\bibinfo {year} {2016}{\natexlab{a}})},\ \Eprint
  {http://arxiv.org/abs/1605.03053} {arXiv:1605.03053} \BibitemShut {NoStop}%
\bibitem [{\citenamefont {{de Martino}}\ \emph
  {et~al.}(2016{\natexlab{b}})\citenamefont {{de Martino}}, \citenamefont
  {{Martins}}, \citenamefont {{Ebeling}},\ and\ \citenamefont
  {{Kocevski}}}]{Martino16b}%
  \BibitemOpen
  \bibfield  {author} {\bibinfo {author} {\bibfnamefont {I.}~\bibnamefont {{de
  Martino}}}, \bibinfo {author} {\bibfnamefont {C.~J.~A.~P.}\ \bibnamefont
  {{Martins}}}, \bibinfo {author} {\bibfnamefont {H.}~\bibnamefont
  {{Ebeling}}}, \ and\ \bibinfo {author} {\bibfnamefont {D.}~\bibnamefont
  {{Kocevski}}},\ }\href {\doibase 10.3390/universe2040034} {\bibfield
  {journal} {\bibinfo  {journal} {Universe}\ }\textbf {\bibinfo {volume} {2}},\
  \bibinfo {eid} {34} (\bibinfo {year} {2016}{\natexlab{b}})},\ \Eprint
  {http://arxiv.org/abs/1612.06739} {arXiv:1612.06739} \BibitemShut {NoStop}%
\bibitem [{\citenamefont {Cola\c{c}o}\ \emph {et~al.}(2019)\citenamefont
  {Cola\c{c}o}, \citenamefont {Holanda}, \citenamefont {Silva},\ and\
  \citenamefont {Alcaniz}}]{Colaco:2019fvl}%
  \BibitemOpen
  \bibfield  {author} {\bibinfo {author} {\bibfnamefont {L.~R.}\ \bibnamefont
  {Cola\c{c}o}}, \bibinfo {author} {\bibfnamefont {R.~F.~L.}\ \bibnamefont
  {Holanda}}, \bibinfo {author} {\bibfnamefont {R.}~\bibnamefont {Silva}}, \
  and\ \bibinfo {author} {\bibfnamefont {J.~S.}\ \bibnamefont {Alcaniz}},\
  }\href {\doibase 10.1088/1475-7516/2019/03/014} {\bibfield  {journal}
  {\bibinfo  {journal} {JCAP}\ }\textbf {\bibinfo {volume} {1903}},\ \bibinfo
  {pages} {014} (\bibinfo {year} {2019})},\ \Eprint
  {http://arxiv.org/abs/1901.10947} {arXiv:1901.10947 [astro-ph.CO]}
  \BibitemShut {NoStop}%
\bibitem [{\citenamefont {{Bergstr{\"o}m}}\ \emph {et~al.}(1999)\citenamefont
  {{Bergstr{\"o}m}}, \citenamefont {{Iguri}},\ and\ \citenamefont
  {{Rubinstein}}}]{bergstrom}%
  \BibitemOpen
  \bibfield  {author} {\bibinfo {author} {\bibfnamefont {L.}~\bibnamefont
  {{Bergstr{\"o}m}}}, \bibinfo {author} {\bibfnamefont {S.}~\bibnamefont
  {{Iguri}}}, \ and\ \bibinfo {author} {\bibfnamefont {H.}~\bibnamefont
  {{Rubinstein}}},\ }\href {\doibase 10.1103/PhysRevD.60.045005} {\bibfield
  {journal} {\bibinfo  {journal} {\prd}\ }\textbf {\bibinfo {volume} {60}},\
  \bibinfo {eid} {045005} (\bibinfo {year} {1999})},\ \Eprint
  {http://arxiv.org/abs/astro-ph/9902157} {astro-ph/9902157} \BibitemShut
  {NoStop}%
\bibitem [{\citenamefont {{Mosquera}}\ and\ \citenamefont
  {{Civitarese}}(2013)}]{mosquera}%
  \BibitemOpen
  \bibfield  {author} {\bibinfo {author} {\bibfnamefont {M.~E.}\ \bibnamefont
  {{Mosquera}}}\ and\ \bibinfo {author} {\bibfnamefont {O.}~\bibnamefont
  {{Civitarese}}},\ }\href {\doibase 10.1051/0004-6361/201220615} {\bibfield
  {journal} {\bibinfo  {journal} {Astronomy and Astrophysics}\ }\textbf
  {\bibinfo {volume} {551}},\ \bibinfo {eid} {A122} (\bibinfo {year}
  {2013})}\BibitemShut {NoStop}%
\bibitem [{\citenamefont {{Battye}}\ \emph {et~al.}(2001)\citenamefont
  {{Battye}}, \citenamefont {{Crittenden}},\ and\ \citenamefont
  {{Weller}}}]{bat}%
  \BibitemOpen
  \bibfield  {author} {\bibinfo {author} {\bibfnamefont {R.~A.}\ \bibnamefont
  {{Battye}}}, \bibinfo {author} {\bibfnamefont {R.}~\bibnamefont
  {{Crittenden}}}, \ and\ \bibinfo {author} {\bibfnamefont {J.}~\bibnamefont
  {{Weller}}},\ }\href {\doibase 10.1103/PhysRevD.63.043505} {\bibfield
  {journal} {\bibinfo  {journal} {\prd}\ }\textbf {\bibinfo {volume} {63}},\
  \bibinfo {eid} {043505} (\bibinfo {year} {2001})},\ \Eprint
  {http://arxiv.org/abs/astro-ph/0008265} {astro-ph/0008265} \BibitemShut
  {NoStop}%
\bibitem [{\citenamefont {{Avelino}}\ \emph {et~al.}(2000)\citenamefont
  {{Avelino}}, \citenamefont {{Martins}}, \citenamefont {{Rocha}},\ and\
  \citenamefont {{Viana}}}]{avelino}%
  \BibitemOpen
  \bibfield  {author} {\bibinfo {author} {\bibfnamefont {P.~P.}\ \bibnamefont
  {{Avelino}}}, \bibinfo {author} {\bibfnamefont {C.~J.~A.~P.}\ \bibnamefont
  {{Martins}}}, \bibinfo {author} {\bibfnamefont {G.}~\bibnamefont {{Rocha}}},
  \ and\ \bibinfo {author} {\bibfnamefont {P.}~\bibnamefont {{Viana}}},\ }\href
  {\doibase 10.1103/PhysRevD.62.123508} {\bibfield  {journal} {\bibinfo
  {journal} {\prd}\ }\textbf {\bibinfo {volume} {62}},\ \bibinfo {eid} {123508}
  (\bibinfo {year} {2000})},\ \Eprint {http://arxiv.org/abs/astro-ph/0008446}
  {astro-ph/0008446} \BibitemShut {NoStop}%
\bibitem [{\citenamefont {{Planck Collaboration}}(2015)}]{Planck2015}%
  \BibitemOpen
  \bibfield  {author} {\bibinfo {author} {\bibnamefont {{Planck
  Collaboration}}},\ }\href {\doibase 10.1051/0004-6361/201424496} {\bibfield
  {journal} {\bibinfo  {journal} {Astronomy and Astrophysics}\ }\textbf
  {\bibinfo {volume} {580}},\ \bibinfo {eid} {A22} (\bibinfo {year} {2015})},\
  \Eprint {http://arxiv.org/abs/1406.7482} {arXiv:1406.7482} \BibitemShut
  {NoStop}%
\bibitem [{\citenamefont {{O'Bryan}}\ \emph {et~al.}(2015)\citenamefont
  {{O'Bryan}}, \citenamefont {{Smidt}}, \citenamefont {{De Bernardis}},\ and\
  \citenamefont {{Cooray}}}]{obryan15}%
  \BibitemOpen
  \bibfield  {author} {\bibinfo {author} {\bibfnamefont {J.}~\bibnamefont
  {{O'Bryan}}}, \bibinfo {author} {\bibfnamefont {J.}~\bibnamefont {{Smidt}}},
  \bibinfo {author} {\bibfnamefont {F.}~\bibnamefont {{De Bernardis}}}, \ and\
  \bibinfo {author} {\bibfnamefont {A.}~\bibnamefont {{Cooray}}},\ }\href
  {\doibase 10.1088/0004-637X/798/1/18} {\bibfield  {journal} {\bibinfo
  {journal} {\apj}\ }\textbf {\bibinfo {volume} {798}},\ \bibinfo {eid} {18}
  (\bibinfo {year} {2015})}\BibitemShut {NoStop}%
\bibitem [{\citenamefont {{Garc{\'{\i}}a-Berro}}\ \emph
  {et~al.}(2007)\citenamefont {{Garc{\'{\i}}a-Berro}}, \citenamefont
  {{Isern}},\ and\ \citenamefont {{Kubyshin}}}]{Yuri}%
  \BibitemOpen
  \bibfield  {author} {\bibinfo {author} {\bibfnamefont {E.}~\bibnamefont
  {{Garc{\'{\i}}a-Berro}}}, \bibinfo {author} {\bibfnamefont {J.}~\bibnamefont
  {{Isern}}}, \ and\ \bibinfo {author} {\bibfnamefont {Y.~A.}\ \bibnamefont
  {{Kubyshin}}},\ }\href {\doibase 10.1007/s00159-006-0004-8} {\bibfield
  {journal} {\bibinfo  {journal} {Astronomy and Astrophysicsr}\ }\textbf
  {\bibinfo {volume} {14}},\ \bibinfo {pages} {113} (\bibinfo {year}
  {2007})}\BibitemShut {NoStop}%
\bibitem [{\citenamefont {{Lor{\'e}n-Aguilar}}\ \emph
  {et~al.}(2003)\citenamefont {{Lor{\'e}n-Aguilar}}, \citenamefont
  {{Garc{\'{\i}}a-Berro}}, \citenamefont {{Isern}},\ and\ \citenamefont
  {{Kubyshin}}}]{Pablo}%
  \BibitemOpen
  \bibfield  {author} {\bibinfo {author} {\bibfnamefont {P.}~\bibnamefont
  {{Lor{\'e}n-Aguilar}}}, \bibinfo {author} {\bibfnamefont {E.}~\bibnamefont
  {{Garc{\'{\i}}a-Berro}}}, \bibinfo {author} {\bibfnamefont {J.}~\bibnamefont
  {{Isern}}}, \ and\ \bibinfo {author} {\bibfnamefont {Y.~A.}\ \bibnamefont
  {{Kubyshin}}},\ }\href {\doibase 10.1088/0264-9381/20/18/302} {\bibfield
  {journal} {\bibinfo  {journal} {Classical and Quantum Gravity}\ }\textbf
  {\bibinfo {volume} {20}},\ \bibinfo {pages} {3885} (\bibinfo {year}
  {2003})},\ \Eprint {http://arxiv.org/abs/astro-ph/0309722} {astro-ph/0309722}
  \BibitemShut {NoStop}%
\bibitem [{\citenamefont {{Bekenstein}}(1982)}]{bekenstein82}%
  \BibitemOpen
  \bibfield  {author} {\bibinfo {author} {\bibfnamefont {J.~D.}\ \bibnamefont
  {{Bekenstein}}},\ }\href {\doibase 10.1103/PhysRevD.25.1527} {\bibfield
  {journal} {\bibinfo  {journal} {\prd}\ }\textbf {\bibinfo {volume} {25}},\
  \bibinfo {pages} {1527} (\bibinfo {year} {1982})}\BibitemShut {NoStop}%
\bibitem [{\citenamefont {{Bekenstein}}(2002)}]{bekenstein2002}%
  \BibitemOpen
  \bibfield  {author} {\bibinfo {author} {\bibfnamefont {J.~D.}\ \bibnamefont
  {{Bekenstein}}},\ }\href {\doibase 10.1103/PhysRevD.66.123514} {\bibfield
  {journal} {\bibinfo  {journal} {\prd}\ }\textbf {\bibinfo {volume} {66}},\
  \bibinfo {eid} {123514} (\bibinfo {year} {2002})},\ \Eprint
  {http://arxiv.org/abs/gr-qc/0208081} {gr-qc/0208081} \BibitemShut {NoStop}%
\bibitem [{\citenamefont {{Barrow}}\ \emph {et~al.}(2002)\citenamefont
  {{Barrow}}, \citenamefont {{Sandvik}},\ and\ \citenamefont
  {{Magueijo}}}]{bsm02}%
  \BibitemOpen
  \bibfield  {author} {\bibinfo {author} {\bibfnamefont {J.~D.}\ \bibnamefont
  {{Barrow}}}, \bibinfo {author} {\bibfnamefont {H.~B.}\ \bibnamefont
  {{Sandvik}}}, \ and\ \bibinfo {author} {\bibfnamefont {J.}~\bibnamefont
  {{Magueijo}}},\ }\href {\doibase 10.1103/PhysRevD.65.063504} {\bibfield
  {journal} {\bibinfo  {journal} {\prd}\ }\textbf {\bibinfo {volume} {65}},\
  \bibinfo {eid} {063504} (\bibinfo {year} {2002})},\ \Eprint
  {http://arxiv.org/abs/astro-ph/0109414} {astro-ph/0109414} \BibitemShut
  {NoStop}%
\bibitem [{\citenamefont {{Barrow}}\ and\ \citenamefont
  {{Magueijo}}(2005)}]{BM05}%
  \BibitemOpen
  \bibfield  {author} {\bibinfo {author} {\bibfnamefont {J.~D.}\ \bibnamefont
  {{Barrow}}}\ and\ \bibinfo {author} {\bibfnamefont {J.}~\bibnamefont
  {{Magueijo}}},\ }\href {\doibase 10.1103/PhysRevD.72.043521} {\bibfield
  {journal} {\bibinfo  {journal} {\prd}\ }\textbf {\bibinfo {volume} {72}},\
  \bibinfo {eid} {043521} (\bibinfo {year} {2005})},\ \Eprint
  {http://arxiv.org/abs/astro-ph/0503222} {astro-ph/0503222} \BibitemShut
  {NoStop}%
\bibitem [{\citenamefont {{Hees}}\ \emph {et~al.}(2014)\citenamefont {{Hees}},
  \citenamefont {{Minazzoli}},\ and\ \citenamefont {{Larena}}}]{Hees2014}%
  \BibitemOpen
  \bibfield  {author} {\bibinfo {author} {\bibfnamefont {A.}~\bibnamefont
  {{Hees}}}, \bibinfo {author} {\bibfnamefont {O.}~\bibnamefont {{Minazzoli}}},
  \ and\ \bibinfo {author} {\bibfnamefont {J.}~\bibnamefont {{Larena}}},\
  }\href {\doibase 10.1103/PhysRevD.90.124064} {\bibfield  {journal} {\bibinfo
  {journal} {\prd}\ }\textbf {\bibinfo {volume} {90}},\ \bibinfo {eid} {124064}
  (\bibinfo {year} {2014})},\ \Eprint {http://arxiv.org/abs/1406.6187}
  {arXiv:1406.6187 [astro-ph.CO]} \BibitemShut {NoStop}%
\bibitem [{\citenamefont {{Minazzoli}}\ and\ \citenamefont
  {{Hees}}(2014)}]{Minazzoli2014}%
  \BibitemOpen
  \bibfield  {author} {\bibinfo {author} {\bibfnamefont {O.}~\bibnamefont
  {{Minazzoli}}}\ and\ \bibinfo {author} {\bibfnamefont {A.}~\bibnamefont
  {{Hees}}},\ }\href {\doibase 10.1103/PhysRevD.90.023017} {\bibfield
  {journal} {\bibinfo  {journal} {\prd}\ }\textbf {\bibinfo {volume} {90}},\
  \bibinfo {eid} {023017} (\bibinfo {year} {2014})},\ \Eprint
  {http://arxiv.org/abs/1404.4266} {arXiv:1404.4266 [gr-qc]} \BibitemShut
  {NoStop}%
\bibitem [{\citenamefont {{Ellis}}(2007)}]{2007GReGr..39.1047E}%
  \BibitemOpen
  \bibfield  {author} {\bibinfo {author} {\bibfnamefont {G.~F.~R.}\
  \bibnamefont {{Ellis}}},\ }\href {\doibase 10.1007/s10714-006-0355-5}
  {\bibfield  {journal} {\bibinfo  {journal} {General Relativity and
  Gravitation}\ }\textbf {\bibinfo {volume} {39}},\ \bibinfo {pages} {1047}
  (\bibinfo {year} {2007})}\BibitemShut {NoStop}%
\bibitem [{\citenamefont {Bassett}\ and\ \citenamefont
  {Kunz}(2004)}]{Bassett:2003vu}%
  \BibitemOpen
  \bibfield  {author} {\bibinfo {author} {\bibfnamefont {B.~A.}\ \bibnamefont
  {Bassett}}\ and\ \bibinfo {author} {\bibfnamefont {M.}~\bibnamefont {Kunz}},\
  }\href {\doibase 10.1103/PhysRevD.69.101305} {\bibfield  {journal} {\bibinfo
  {journal} {Phys. Rev.}\ }\textbf {\bibinfo {volume} {D69}},\ \bibinfo {pages}
  {101305} (\bibinfo {year} {2004})},\ \Eprint
  {http://arxiv.org/abs/astro-ph/0312443} {arXiv:astro-ph/0312443 [astro-ph]}
  \BibitemShut {NoStop}%
\bibitem [{\citenamefont {Gon\c{c}alves}\ \emph {et~al.}(2012)\citenamefont
  {Gon\c{c}alves}, \citenamefont {Holanda},\ and\ \citenamefont
  {Alcaniz}}]{Goncalves:2011ha}%
  \BibitemOpen
  \bibfield  {author} {\bibinfo {author} {\bibfnamefont {R.~S.}\ \bibnamefont
  {Gon\c{c}alves}}, \bibinfo {author} {\bibfnamefont {R.~F.~L.}\ \bibnamefont
  {Holanda}}, \ and\ \bibinfo {author} {\bibfnamefont {J.~S.}\ \bibnamefont
  {Alcaniz}},\ }\href {\doibase 10.1111/j.1745-3933.2011.01192.x} {\bibfield
  {journal} {\bibinfo  {journal} {Mon. Not. Roy. Astron. Soc.}\ }\textbf
  {\bibinfo {volume} {420}},\ \bibinfo {pages} {L43} (\bibinfo {year}
  {2012})},\ \Eprint {http://arxiv.org/abs/1109.2790} {arXiv:1109.2790
  [astro-ph.CO]} \BibitemShut {NoStop}%
\bibitem [{\citenamefont {Holanda}\ \emph {et~al.}(2012)\citenamefont
  {Holanda}, \citenamefont {Gon\c{c}alves},\ and\ \citenamefont
  {Alcaniz}}]{Holanda:2012at}%
  \BibitemOpen
  \bibfield  {author} {\bibinfo {author} {\bibfnamefont {R.~F.~L.}\
  \bibnamefont {Holanda}}, \bibinfo {author} {\bibfnamefont {R.~S.}\
  \bibnamefont {Gon\c{c}alves}}, \ and\ \bibinfo {author} {\bibfnamefont
  {J.~S.}\ \bibnamefont {Alcaniz}},\ }\href {\doibase
  10.1088/1475-7516/2012/06/022} {\bibfield  {journal} {\bibinfo  {journal}
  {JCAP}\ }\textbf {\bibinfo {volume} {1206}},\ \bibinfo {pages} {022}
  (\bibinfo {year} {2012})},\ \Eprint {http://arxiv.org/abs/1201.2378}
  {arXiv:1201.2378 [astro-ph.CO]} \BibitemShut {NoStop}%
\bibitem [{\citenamefont {Santana}\ \emph {et~al.}(2017)\citenamefont
  {Santana}, \citenamefont {Calvão}, \citenamefont {Reis},\ and\ \citenamefont
  {Siffert}}]{Santana:2017zvy}%
  \BibitemOpen
  \bibfield  {author} {\bibinfo {author} {\bibfnamefont {L.~T.}\ \bibnamefont
  {Santana}}, \bibinfo {author} {\bibfnamefont {M.~O.}\ \bibnamefont
  {Calvão}}, \bibinfo {author} {\bibfnamefont {R.~R.~R.}\ \bibnamefont
  {Reis}}, \ and\ \bibinfo {author} {\bibfnamefont {B.~B.}\ \bibnamefont
  {Siffert}},\ }\href {\doibase 10.1103/PhysRevD.95.061501} {\bibfield
  {journal} {\bibinfo  {journal} {Phys. Rev.}\ }\textbf {\bibinfo {volume}
  {D95}},\ \bibinfo {pages} {061501} (\bibinfo {year} {2017})},\ \Eprint
  {http://arxiv.org/abs/1703.10871} {arXiv:1703.10871 [gr-qc]} \BibitemShut
  {NoStop}%
\bibitem [{\citenamefont {Ellis}\ \emph {et~al.}(2013)\citenamefont {Ellis},
  \citenamefont {Poltis}, \citenamefont {Uzan},\ and\ \citenamefont
  {Weltman}}]{Ellis:2013cu}%
  \BibitemOpen
  \bibfield  {author} {\bibinfo {author} {\bibfnamefont {G.~F.~R.}\
  \bibnamefont {Ellis}}, \bibinfo {author} {\bibfnamefont {R.}~\bibnamefont
  {Poltis}}, \bibinfo {author} {\bibfnamefont {J.-P.}\ \bibnamefont {Uzan}}, \
  and\ \bibinfo {author} {\bibfnamefont {A.}~\bibnamefont {Weltman}},\ }\href
  {\doibase 10.1103/PhysRevD.87.103530} {\bibfield  {journal} {\bibinfo
  {journal} {Phys. Rev.}\ }\textbf {\bibinfo {volume} {D87}},\ \bibinfo {pages}
  {103530} (\bibinfo {year} {2013})},\ \Eprint {http://arxiv.org/abs/1301.1312}
  {arXiv:1301.1312 [astro-ph.CO]} \BibitemShut {NoStop}%
\bibitem [{\citenamefont {{Ruan}}\ \emph {et~al.}(2018)\citenamefont {{Ruan}},
  \citenamefont {{Melia}},\ and\ \citenamefont {{Zhang}}}]{Ruan:2018}%
  \BibitemOpen
  \bibfield  {author} {\bibinfo {author} {\bibfnamefont {C.-Z.}\ \bibnamefont
  {{Ruan}}}, \bibinfo {author} {\bibfnamefont {F.}~\bibnamefont {{Melia}}}, \
  and\ \bibinfo {author} {\bibfnamefont {T.-J.}\ \bibnamefont {{Zhang}}},\
  }\href {\doibase 10.3847/1538-4357/aaddfd} {\bibfield  {journal} {\bibinfo
  {journal} {\apj}\ }\textbf {\bibinfo {volume} {866}},\ \bibinfo {eid} {31}
  (\bibinfo {year} {2018})},\ \Eprint {http://arxiv.org/abs/1808.09331}
  {arXiv:1808.09331 [astro-ph.CO]} \BibitemShut {NoStop}%
\bibitem [{\citenamefont {{Lin}}\ \emph {et~al.}(2018)\citenamefont {{Lin}},
  \citenamefont {{Li}},\ and\ \citenamefont {{Li}}}]{Lin:2018}%
  \BibitemOpen
  \bibfield  {author} {\bibinfo {author} {\bibfnamefont {H.-N.}\ \bibnamefont
  {{Lin}}}, \bibinfo {author} {\bibfnamefont {M.-H.}\ \bibnamefont {{Li}}}, \
  and\ \bibinfo {author} {\bibfnamefont {X.}~\bibnamefont {{Li}}},\ }\href
  {\doibase 10.1093/mnras/sty2062} {\bibfield  {journal} {\bibinfo  {journal}
  {Monthly Notices Royal Astronomical Society}\ }\textbf {\bibinfo {volume}
  {480}},\ \bibinfo {pages} {3117} (\bibinfo {year} {2018})},\ \Eprint
  {http://arxiv.org/abs/1808.01784} {arXiv:1808.01784 [astro-ph.CO]}
  \BibitemShut {NoStop}%
\bibitem [{\citenamefont {{Ma}}\ and\ \citenamefont
  {{Corasaniti}}(2018)}]{Ma:2018}%
  \BibitemOpen
  \bibfield  {author} {\bibinfo {author} {\bibfnamefont {C.}~\bibnamefont
  {{Ma}}}\ and\ \bibinfo {author} {\bibfnamefont {P.-S.}\ \bibnamefont
  {{Corasaniti}}},\ }\href {\doibase 10.3847/1538-4357/aac88f} {\bibfield
  {journal} {\bibinfo  {journal} {\apj}\ }\textbf {\bibinfo {volume} {861}},\
  \bibinfo {eid} {124} (\bibinfo {year} {2018})},\ \Eprint
  {http://arxiv.org/abs/1604.04631} {arXiv:1604.04631 [astro-ph.CO]}
  \BibitemShut {NoStop}%
\bibitem [{\citenamefont {{Santos-da-Costa}}\ \emph {et~al.}(2015)\citenamefont
  {{Santos-da-Costa}}, \citenamefont {{Busti}},\ and\ \citenamefont
  {{Holanda}}}]{daCosta:2015}%
  \BibitemOpen
  \bibfield  {author} {\bibinfo {author} {\bibfnamefont {S.}~\bibnamefont
  {{Santos-da-Costa}}}, \bibinfo {author} {\bibfnamefont {V.~C.}\ \bibnamefont
  {{Busti}}}, \ and\ \bibinfo {author} {\bibfnamefont {R.~F.~L.}\ \bibnamefont
  {{Holanda}}},\ }\href {\doibase 10.1088/1475-7516/2015/10/061} {\bibfield
  {journal} {\bibinfo  {journal} {Journal of Cosmology and Particle Physics}\
  }\textbf {\bibinfo {volume} {2015}},\ \bibinfo {eid} {061} (\bibinfo {year}
  {2015})},\ \Eprint {http://arxiv.org/abs/1506.00145} {arXiv:1506.00145
  [astro-ph.CO]} \BibitemShut {NoStop}%
\bibitem [{\citenamefont {{Srianand}}\ \emph {et~al.}(2004)\citenamefont
  {{Srianand}}, \citenamefont {{Chand}}, \citenamefont {{Petitjean}},\ and\
  \citenamefont {{Aracil}}}]{srianand}%
  \BibitemOpen
  \bibfield  {author} {\bibinfo {author} {\bibfnamefont {R.}~\bibnamefont
  {{Srianand}}}, \bibinfo {author} {\bibfnamefont {H.}~\bibnamefont {{Chand}}},
  \bibinfo {author} {\bibfnamefont {P.}~\bibnamefont {{Petitjean}}}, \ and\
  \bibinfo {author} {\bibfnamefont {B.}~\bibnamefont {{Aracil}}},\ }\href
  {\doibase 10.1103/PhysRevLett.92.121302} {\bibfield  {journal} {\bibinfo
  {journal} {Physical Review Letters}\ }\textbf {\bibinfo {volume} {92}},\
  \bibinfo {eid} {121302} (\bibinfo {year} {2004})},\ \Eprint
  {http://arxiv.org/abs/astro-ph/0402177} {astro-ph/0402177} \BibitemShut
  {NoStop}%
\bibitem [{\citenamefont {{Webb}}\ \emph {et~al.}(2011)\citenamefont {{Webb}},
  \citenamefont {{King}}, \citenamefont {{Murphy}}, \citenamefont {{Flambaum}},
  \citenamefont {{Carswell}},\ and\ \citenamefont {{Bainbridge}}}]{webb11}%
  \BibitemOpen
  \bibfield  {author} {\bibinfo {author} {\bibfnamefont {J.~K.}\ \bibnamefont
  {{Webb}}}, \bibinfo {author} {\bibfnamefont {J.~A.}\ \bibnamefont {{King}}},
  \bibinfo {author} {\bibfnamefont {M.~T.}\ \bibnamefont {{Murphy}}}, \bibinfo
  {author} {\bibfnamefont {V.~V.}\ \bibnamefont {{Flambaum}}}, \bibinfo
  {author} {\bibfnamefont {R.~F.}\ \bibnamefont {{Carswell}}}, \ and\ \bibinfo
  {author} {\bibfnamefont {M.~B.}\ \bibnamefont {{Bainbridge}}},\ }\href
  {\doibase 10.1103/PhysRevLett.107.191101} {\bibfield  {journal} {\bibinfo
  {journal} {Physical Review Letters}\ }\textbf {\bibinfo {volume} {107}},\
  \bibinfo {eid} {191101} (\bibinfo {year} {2011})},\ \Eprint
  {http://arxiv.org/abs/1008.3907} {arXiv:1008.3907 [astro-ph.CO]} \BibitemShut
  {NoStop}%
\bibitem [{\citenamefont {{King}}\ \emph {et~al.}(2012)\citenamefont {{King}},
  \citenamefont {{Webb}}, \citenamefont {{Murphy}}, \citenamefont {{Flambaum}},
  \citenamefont {{Carswell}}, \citenamefont {{Bainbridge}}, \citenamefont
  {{Wilczynska}},\ and\ \citenamefont {{Koch}}}]{King12}%
  \BibitemOpen
  \bibfield  {author} {\bibinfo {author} {\bibfnamefont {J.~A.}\ \bibnamefont
  {{King}}}, \bibinfo {author} {\bibfnamefont {J.~K.}\ \bibnamefont {{Webb}}},
  \bibinfo {author} {\bibfnamefont {M.~T.}\ \bibnamefont {{Murphy}}}, \bibinfo
  {author} {\bibfnamefont {V.~V.}\ \bibnamefont {{Flambaum}}}, \bibinfo
  {author} {\bibfnamefont {R.~F.}\ \bibnamefont {{Carswell}}}, \bibinfo
  {author} {\bibfnamefont {M.~B.}\ \bibnamefont {{Bainbridge}}}, \bibinfo
  {author} {\bibfnamefont {M.~R.}\ \bibnamefont {{Wilczynska}}}, \ and\
  \bibinfo {author} {\bibfnamefont {F.~E.}\ \bibnamefont {{Koch}}},\ }\href
  {\doibase 10.1111/j.1365-2966.2012.20852.x} {\bibfield  {journal} {\bibinfo
  {journal} {Monthly Notices Royal Astronomical Society}\ }\textbf {\bibinfo
  {volume} {422}},\ \bibinfo {pages} {3370} (\bibinfo {year} {2012})},\ \Eprint
  {http://arxiv.org/abs/1202.4758} {arXiv:1202.4758 [astro-ph.CO]} \BibitemShut
  {NoStop}%
\bibitem [{\citenamefont {{Whitmore}}\ and\ \citenamefont
  {{Murphy}}(2015)}]{whit}%
  \BibitemOpen
  \bibfield  {author} {\bibinfo {author} {\bibfnamefont {J.~B.}\ \bibnamefont
  {{Whitmore}}}\ and\ \bibinfo {author} {\bibfnamefont {M.~T.}\ \bibnamefont
  {{Murphy}}},\ }\href {\doibase 10.1093/mnras/stu2420} {\bibfield  {journal}
  {\bibinfo  {journal} {Monthly Notices Royal Astronomical Society}\ }\textbf
  {\bibinfo {volume} {447}},\ \bibinfo {pages} {446} (\bibinfo {year}
  {2015})},\ \Eprint {http://arxiv.org/abs/1409.4467} {arXiv:1409.4467
  [astro-ph.IM]} \BibitemShut {NoStop}%
\bibitem [{\citenamefont {{Molaro}}\ \emph {et~al.}(2013)\citenamefont
  {{Molaro}}, \citenamefont {{Centuri{\'o}n}}, \citenamefont {{Whitmore}},
  \citenamefont {{Evans}}, \citenamefont {{Murphy}}, \citenamefont
  {{Agafonova}}, \citenamefont {{Bonifacio}}, \citenamefont {{D'Odorico}},
  \citenamefont {{Levshakov}}, \citenamefont {{Lopez}}, \citenamefont
  {{Martins}}, \citenamefont {{Petitjean}}, \citenamefont {{Rahmani}},
  \citenamefont {{Reimers}}, \citenamefont {{Srianand}}, \citenamefont
  {{Vladilo}},\ and\ \citenamefont {{Wendt}}}]{Molaro2013}%
  \BibitemOpen
  \bibfield  {author} {\bibinfo {author} {\bibfnamefont {P.}~\bibnamefont
  {{Molaro}}}, \bibinfo {author} {\bibfnamefont {M.}~\bibnamefont
  {{Centuri{\'o}n}}}, \bibinfo {author} {\bibfnamefont {J.~B.}\ \bibnamefont
  {{Whitmore}}}, \bibinfo {author} {\bibfnamefont {T.~M.}\ \bibnamefont
  {{Evans}}}, \bibinfo {author} {\bibfnamefont {M.~T.}\ \bibnamefont
  {{Murphy}}}, \bibinfo {author} {\bibfnamefont {I.~I.}\ \bibnamefont
  {{Agafonova}}}, \bibinfo {author} {\bibfnamefont {P.}~\bibnamefont
  {{Bonifacio}}}, \bibinfo {author} {\bibfnamefont {S.}~\bibnamefont
  {{D'Odorico}}}, \bibinfo {author} {\bibfnamefont {S.~A.}\ \bibnamefont
  {{Levshakov}}}, \bibinfo {author} {\bibfnamefont {S.}~\bibnamefont
  {{Lopez}}}, \bibinfo {author} {\bibfnamefont {C.~J.~A.~P.}\ \bibnamefont
  {{Martins}}}, \bibinfo {author} {\bibfnamefont {P.}~\bibnamefont
  {{Petitjean}}}, \bibinfo {author} {\bibfnamefont {H.}~\bibnamefont
  {{Rahmani}}}, \bibinfo {author} {\bibfnamefont {D.}~\bibnamefont
  {{Reimers}}}, \bibinfo {author} {\bibfnamefont {R.}~\bibnamefont
  {{Srianand}}}, \bibinfo {author} {\bibfnamefont {G.}~\bibnamefont
  {{Vladilo}}}, \ and\ \bibinfo {author} {\bibfnamefont {M.}~\bibnamefont
  {{Wendt}}},\ }\href {\doibase 10.1051/0004-6361/201321351} {\bibfield
  {journal} {\bibinfo  {journal} {Astronomy and Astrophysics}\ }\textbf
  {\bibinfo {volume} {555}},\ \bibinfo {eid} {A68} (\bibinfo {year} {2013})},\
  \Eprint {http://arxiv.org/abs/1305.1884} {arXiv:1305.1884 [astro-ph.CO]}
  \BibitemShut {NoStop}%
\bibitem [{\citenamefont {{Rahmani}}\ \emph {et~al.}(2013)\citenamefont
  {{Rahmani}}, \citenamefont {{Wendt}}, \citenamefont {{Srianand}},
  \citenamefont {{Noterdaeme}}, \citenamefont {{Petitjean}}, \citenamefont
  {{Molaro}}, \citenamefont {{Whitmore}}, \citenamefont {{Murphy}},
  \citenamefont {{Centurion}}, \citenamefont {{Fathivavsari}}, \citenamefont
  {{D'Odorico}}, \citenamefont {{Evans}}, \citenamefont {{Levshakov}},
  \citenamefont {{Lopez}}, \citenamefont {{Martins}}, \citenamefont
  {{Reimers}},\ and\ \citenamefont {{Vladilo}}}]{Rahmani2013}%
  \BibitemOpen
  \bibfield  {author} {\bibinfo {author} {\bibfnamefont {H.}~\bibnamefont
  {{Rahmani}}}, \bibinfo {author} {\bibfnamefont {M.}~\bibnamefont {{Wendt}}},
  \bibinfo {author} {\bibfnamefont {R.}~\bibnamefont {{Srianand}}}, \bibinfo
  {author} {\bibfnamefont {P.}~\bibnamefont {{Noterdaeme}}}, \bibinfo {author}
  {\bibfnamefont {P.}~\bibnamefont {{Petitjean}}}, \bibinfo {author}
  {\bibfnamefont {P.}~\bibnamefont {{Molaro}}}, \bibinfo {author}
  {\bibfnamefont {J.~B.}\ \bibnamefont {{Whitmore}}}, \bibinfo {author}
  {\bibfnamefont {M.~T.}\ \bibnamefont {{Murphy}}}, \bibinfo {author}
  {\bibfnamefont {M.}~\bibnamefont {{Centurion}}}, \bibinfo {author}
  {\bibfnamefont {H.}~\bibnamefont {{Fathivavsari}}}, \bibinfo {author}
  {\bibfnamefont {S.}~\bibnamefont {{D'Odorico}}}, \bibinfo {author}
  {\bibfnamefont {T.~M.}\ \bibnamefont {{Evans}}}, \bibinfo {author}
  {\bibfnamefont {S.~A.}\ \bibnamefont {{Levshakov}}}, \bibinfo {author}
  {\bibfnamefont {S.}~\bibnamefont {{Lopez}}}, \bibinfo {author} {\bibfnamefont
  {C.~J.~A.~P.}\ \bibnamefont {{Martins}}}, \bibinfo {author} {\bibfnamefont
  {D.}~\bibnamefont {{Reimers}}}, \ and\ \bibinfo {author} {\bibfnamefont
  {G.}~\bibnamefont {{Vladilo}}},\ }\href {\doibase 10.1093/mnras/stt1356}
  {\bibfield  {journal} {\bibinfo  {journal} {Monthly Notices Royal
  Astronomical Society}\ }\textbf {\bibinfo {volume} {435}},\ \bibinfo {pages}
  {861} (\bibinfo {year} {2013})},\ \Eprint {http://arxiv.org/abs/1307.5864}
  {arXiv:1307.5864 [astro-ph.CO]} \BibitemShut {NoStop}%
\bibitem [{\citenamefont {{Evans}}\ \emph {et~al.}(2014)\citenamefont
  {{Evans}}, \citenamefont {{Murphy}}, \citenamefont {{Whitmore}},
  \citenamefont {{Misawa}}, \citenamefont {{Centurion}}, \citenamefont
  {{D'Odorico}}, \citenamefont {{Lopez}}, \citenamefont {{Martins}},
  \citenamefont {{Molaro}}, \citenamefont {{Petitjean}}, \citenamefont
  {{Rahmani}}, \citenamefont {{Srianand}},\ and\ \citenamefont
  {{Wendt}}}]{Evans2014}%
  \BibitemOpen
  \bibfield  {author} {\bibinfo {author} {\bibfnamefont {T.~M.}\ \bibnamefont
  {{Evans}}}, \bibinfo {author} {\bibfnamefont {M.~T.}\ \bibnamefont
  {{Murphy}}}, \bibinfo {author} {\bibfnamefont {J.~B.}\ \bibnamefont
  {{Whitmore}}}, \bibinfo {author} {\bibfnamefont {T.}~\bibnamefont
  {{Misawa}}}, \bibinfo {author} {\bibfnamefont {M.}~\bibnamefont
  {{Centurion}}}, \bibinfo {author} {\bibfnamefont {S.}~\bibnamefont
  {{D'Odorico}}}, \bibinfo {author} {\bibfnamefont {S.}~\bibnamefont
  {{Lopez}}}, \bibinfo {author} {\bibfnamefont {C.~J.~A.~P.}\ \bibnamefont
  {{Martins}}}, \bibinfo {author} {\bibfnamefont {P.}~\bibnamefont {{Molaro}}},
  \bibinfo {author} {\bibfnamefont {P.}~\bibnamefont {{Petitjean}}}, \bibinfo
  {author} {\bibfnamefont {H.}~\bibnamefont {{Rahmani}}}, \bibinfo {author}
  {\bibfnamefont {R.}~\bibnamefont {{Srianand}}}, \ and\ \bibinfo {author}
  {\bibfnamefont {M.}~\bibnamefont {{Wendt}}},\ }\href {\doibase
  10.1093/mnras/stu1754} {\bibfield  {journal} {\bibinfo  {journal} {Monthly
  Notices Royal Astronomical Society}\ }\textbf {\bibinfo {volume} {445}},\
  \bibinfo {pages} {128} (\bibinfo {year} {2014})},\ \Eprint
  {http://arxiv.org/abs/1409.1923} {arXiv:1409.1923 [astro-ph.CO]} \BibitemShut
  {NoStop}%
\bibitem [{\citenamefont {{Pepe}}\ \emph {et~al.}(2013)\citenamefont {{Pepe}},
  \citenamefont {{Cristiani}}, \citenamefont {{Rebolo}}, \citenamefont
  {{Santos}}, \citenamefont {{Dekker}}, \citenamefont {{M{\'e}gevand}},
  \citenamefont {{Zerbi}}, \citenamefont {{Cabral}}, \citenamefont {{Molaro}},
  \citenamefont {{Di Marcantonio}}, \citenamefont {{Abreu}}, \citenamefont
  {{Affolter}}, \citenamefont {{Aliverti}}, \citenamefont {{Allende Prieto}},
  \citenamefont {{Amate}}, \citenamefont {{Avila}}, \citenamefont {{Baldini}},
  \citenamefont {{Bristow}}, \citenamefont {{Broeg}}, \citenamefont {{Cirami}},
  \citenamefont {{Coelho}}, \citenamefont {{Conconi}}, \citenamefont
  {{Coretti}}, \citenamefont {{Cupani}}, \citenamefont {{D'Odorico}},
  \citenamefont {{De Caprio}}, \citenamefont {{Delabre}}, \citenamefont
  {{Dorn}}, \citenamefont {{Figueira}}, \citenamefont {{Fragoso}},
  \citenamefont {{Galeotta}}, \citenamefont {{Genolet}}, \citenamefont
  {{Gomes}}, \citenamefont {{Gonz{\'a}lez Hern{\'a}ndez}}, \citenamefont
  {{Hughes}}, \citenamefont {{Iwert}}, \citenamefont {{Kerber}}, \citenamefont
  {{Landoni}}, \citenamefont {{Lizon}}, \citenamefont {{Lovis}}, \citenamefont
  {{Maire}}, \citenamefont {{Mannetta}}, \citenamefont {{Martins}},
  \citenamefont {{Monteiro}}, \citenamefont {{Oliveira}}, \citenamefont
  {{Poretti}}, \citenamefont {{Rasilla}}, \citenamefont {{Riva}}, \citenamefont
  {{Santana Tschudi}}, \citenamefont {{Santos}}, \citenamefont {{Sosnowska}},
  \citenamefont {{Sousa}}, \citenamefont {{Span{\`o}}}, \citenamefont
  {{Tenegi}}, \citenamefont {{Toso}}, \citenamefont {{Vanzella}}, \citenamefont
  {{Viel}},\ and\ \citenamefont {{Zapatero Osorio}}}]{ESPRESSO}%
  \BibitemOpen
  \bibfield  {author} {\bibinfo {author} {\bibfnamefont {F.}~\bibnamefont
  {{Pepe}}}, \bibinfo {author} {\bibfnamefont {S.}~\bibnamefont {{Cristiani}}},
  \bibinfo {author} {\bibfnamefont {R.}~\bibnamefont {{Rebolo}}}, \bibinfo
  {author} {\bibfnamefont {N.~C.}\ \bibnamefont {{Santos}}}, \bibinfo {author}
  {\bibfnamefont {H.}~\bibnamefont {{Dekker}}}, \bibinfo {author}
  {\bibfnamefont {D.}~\bibnamefont {{M{\'e}gevand}}}, \bibinfo {author}
  {\bibfnamefont {F.~M.}\ \bibnamefont {{Zerbi}}}, \bibinfo {author}
  {\bibfnamefont {A.}~\bibnamefont {{Cabral}}}, \bibinfo {author}
  {\bibfnamefont {P.}~\bibnamefont {{Molaro}}}, \bibinfo {author}
  {\bibfnamefont {P.}~\bibnamefont {{Di Marcantonio}}}, \bibinfo {author}
  {\bibfnamefont {M.}~\bibnamefont {{Abreu}}}, \bibinfo {author} {\bibfnamefont
  {M.}~\bibnamefont {{Affolter}}}, \bibinfo {author} {\bibfnamefont
  {M.}~\bibnamefont {{Aliverti}}}, \bibinfo {author} {\bibfnamefont
  {C.}~\bibnamefont {{Allende Prieto}}}, \bibinfo {author} {\bibfnamefont
  {M.}~\bibnamefont {{Amate}}}, \bibinfo {author} {\bibfnamefont
  {G.}~\bibnamefont {{Avila}}}, \bibinfo {author} {\bibfnamefont
  {V.}~\bibnamefont {{Baldini}}}, \bibinfo {author} {\bibfnamefont
  {P.}~\bibnamefont {{Bristow}}}, \bibinfo {author} {\bibfnamefont
  {C.}~\bibnamefont {{Broeg}}}, \bibinfo {author} {\bibfnamefont
  {R.}~\bibnamefont {{Cirami}}}, \bibinfo {author} {\bibfnamefont
  {J.}~\bibnamefont {{Coelho}}}, \bibinfo {author} {\bibfnamefont
  {P.}~\bibnamefont {{Conconi}}}, \bibinfo {author} {\bibfnamefont
  {I.}~\bibnamefont {{Coretti}}}, \bibinfo {author} {\bibfnamefont
  {G.}~\bibnamefont {{Cupani}}}, \bibinfo {author} {\bibfnamefont
  {V.}~\bibnamefont {{D'Odorico}}}, \bibinfo {author} {\bibfnamefont
  {V.}~\bibnamefont {{De Caprio}}}, \bibinfo {author} {\bibfnamefont
  {B.}~\bibnamefont {{Delabre}}}, \bibinfo {author} {\bibfnamefont
  {R.}~\bibnamefont {{Dorn}}}, \bibinfo {author} {\bibfnamefont
  {P.}~\bibnamefont {{Figueira}}}, \bibinfo {author} {\bibfnamefont
  {A.}~\bibnamefont {{Fragoso}}}, \bibinfo {author} {\bibfnamefont
  {S.}~\bibnamefont {{Galeotta}}}, \bibinfo {author} {\bibfnamefont
  {L.}~\bibnamefont {{Genolet}}}, \bibinfo {author} {\bibfnamefont
  {R.}~\bibnamefont {{Gomes}}}, \bibinfo {author} {\bibfnamefont {J.~I.}\
  \bibnamefont {{Gonz{\'a}lez Hern{\'a}ndez}}}, \bibinfo {author}
  {\bibfnamefont {I.}~\bibnamefont {{Hughes}}}, \bibinfo {author}
  {\bibfnamefont {O.}~\bibnamefont {{Iwert}}}, \bibinfo {author} {\bibfnamefont
  {F.}~\bibnamefont {{Kerber}}}, \bibinfo {author} {\bibfnamefont
  {M.}~\bibnamefont {{Landoni}}}, \bibinfo {author} {\bibfnamefont {J.~L.}\
  \bibnamefont {{Lizon}}}, \bibinfo {author} {\bibfnamefont {C.}~\bibnamefont
  {{Lovis}}}, \bibinfo {author} {\bibfnamefont {C.}~\bibnamefont {{Maire}}},
  \bibinfo {author} {\bibfnamefont {M.}~\bibnamefont {{Mannetta}}}, \bibinfo
  {author} {\bibfnamefont {C.}~\bibnamefont {{Martins}}}, \bibinfo {author}
  {\bibfnamefont {M.~A.}\ \bibnamefont {{Monteiro}}}, \bibinfo {author}
  {\bibfnamefont {A.}~\bibnamefont {{Oliveira}}}, \bibinfo {author}
  {\bibfnamefont {E.}~\bibnamefont {{Poretti}}}, \bibinfo {author}
  {\bibfnamefont {J.~L.}\ \bibnamefont {{Rasilla}}}, \bibinfo {author}
  {\bibfnamefont {M.}~\bibnamefont {{Riva}}}, \bibinfo {author} {\bibfnamefont
  {S.}~\bibnamefont {{Santana Tschudi}}}, \bibinfo {author} {\bibfnamefont
  {P.}~\bibnamefont {{Santos}}}, \bibinfo {author} {\bibfnamefont
  {D.}~\bibnamefont {{Sosnowska}}}, \bibinfo {author} {\bibfnamefont
  {S.}~\bibnamefont {{Sousa}}}, \bibinfo {author} {\bibfnamefont
  {P.}~\bibnamefont {{Span{\`o}}}}, \bibinfo {author} {\bibfnamefont
  {F.}~\bibnamefont {{Tenegi}}}, \bibinfo {author} {\bibfnamefont
  {G.}~\bibnamefont {{Toso}}}, \bibinfo {author} {\bibfnamefont
  {E.}~\bibnamefont {{Vanzella}}}, \bibinfo {author} {\bibfnamefont
  {M.}~\bibnamefont {{Viel}}}, \ and\ \bibinfo {author} {\bibfnamefont {M.~R.}\
  \bibnamefont {{Zapatero Osorio}}},\ }\href@noop {} {\bibfield  {journal}
  {\bibinfo  {journal} {The Messenger}\ }\textbf {\bibinfo {volume} {153}},\
  \bibinfo {pages} {6} (\bibinfo {year} {2013})}\BibitemShut {NoStop}%
\bibitem [{\citenamefont {{Marconi}}\ \emph {et~al.}(2018)\citenamefont
  {{Marconi}}, \citenamefont {{Allende Prieto}}, \citenamefont {{Amado}},
  \citenamefont {{Amate}}, \citenamefont {{Augusto}}, \citenamefont
  {{Becerril}}, \citenamefont {{Bezawada}}, \citenamefont {{Boisse}},
  \citenamefont {{Bouchy}}, \citenamefont {{Cabral}}, \citenamefont
  {{Chazelas}}, \citenamefont {{Cirami}}, \citenamefont {{Coretti}},
  \citenamefont {{Cristiani}}, \citenamefont {{Cupani}}, \citenamefont {{de
  Castro Le{\~a}o}}, \citenamefont {{de Medeiros}}, \citenamefont {{de Souza}},
  \citenamefont {{Di Marcantonio}}, \citenamefont {{Di Varano}}, \citenamefont
  {{D'Odorico}}, \citenamefont {{Drass}}, \citenamefont {{Figueira}},
  \citenamefont {{Fragoso}}, \citenamefont {{Fynbo}}, \citenamefont {{Genoni}},
  \citenamefont {{Gonz{\'a}lez Hern{\'a}ndez}}, \citenamefont {{Haehnelt}},
  \citenamefont {{Hughes}}, \citenamefont {{Huke}}, \citenamefont {{Kjeldsen}},
  \citenamefont {{Korn}}, \citenamefont {{Land oni}}, \citenamefont {{Liske}},
  \citenamefont {{Lovis}}, \citenamefont {{Maiolino}}, \citenamefont
  {{Marquart}}, \citenamefont {{Martins}}, \citenamefont {{Mason}},
  \citenamefont {{Monteiro}}, \citenamefont {{Morris}}, \citenamefont
  {{Murray}}, \citenamefont {{Niedzielski}}, \citenamefont {{Oliva}},
  \citenamefont {{Origlia}}, \citenamefont {{Pall{\'e}}}, \citenamefont
  {{Parr-Burman}}, \citenamefont {{Parro}}, \citenamefont {{Pepe}},
  \citenamefont {{Piskunov}}, \citenamefont {{Rasilla}}, \citenamefont
  {{Rees}}, \citenamefont {{Rebolo}}, \citenamefont {{Riva}}, \citenamefont
  {{Rousseau}}, \citenamefont {{Sanna}}, \citenamefont {{Santos}},
  \citenamefont {{Shen}}, \citenamefont {{Sortino}}, \citenamefont
  {{Sosnowska}}, \citenamefont {{Sousa}}, \citenamefont {{Stempels}},
  \citenamefont {{Strassmeier}}, \citenamefont {{Tenegi}}, \citenamefont
  {{Tozzi}}, \citenamefont {{Udry}}, \citenamefont {{Valenziano}},
  \citenamefont {{Vanzi}}, \citenamefont {{Weber}}, \citenamefont {{Woche}},
  \citenamefont {{Xompero}},\ and\ \citenamefont {{Zackrisson}}}]{ELT}%
  \BibitemOpen
  \bibfield  {author} {\bibinfo {author} {\bibfnamefont {A.}~\bibnamefont
  {{Marconi}}}, \bibinfo {author} {\bibfnamefont {C.}~\bibnamefont {{Allende
  Prieto}}}, \bibinfo {author} {\bibfnamefont {P.~J.}\ \bibnamefont {{Amado}}},
  \bibinfo {author} {\bibfnamefont {M.}~\bibnamefont {{Amate}}}, \bibinfo
  {author} {\bibfnamefont {S.~R.}\ \bibnamefont {{Augusto}}}, \bibinfo {author}
  {\bibfnamefont {S.}~\bibnamefont {{Becerril}}}, \bibinfo {author}
  {\bibfnamefont {N.}~\bibnamefont {{Bezawada}}}, \bibinfo {author}
  {\bibfnamefont {I.}~\bibnamefont {{Boisse}}}, \bibinfo {author}
  {\bibfnamefont {F.}~\bibnamefont {{Bouchy}}}, \bibinfo {author}
  {\bibfnamefont {A.}~\bibnamefont {{Cabral}}}, \bibinfo {author}
  {\bibfnamefont {B.}~\bibnamefont {{Chazelas}}}, \bibinfo {author}
  {\bibfnamefont {R.}~\bibnamefont {{Cirami}}}, \bibinfo {author}
  {\bibfnamefont {I.}~\bibnamefont {{Coretti}}}, \bibinfo {author}
  {\bibfnamefont {S.}~\bibnamefont {{Cristiani}}}, \bibinfo {author}
  {\bibfnamefont {G.}~\bibnamefont {{Cupani}}}, \bibinfo {author}
  {\bibfnamefont {I.}~\bibnamefont {{de Castro Le{\~a}o}}}, \bibinfo {author}
  {\bibfnamefont {J.~R.}\ \bibnamefont {{de Medeiros}}}, \bibinfo {author}
  {\bibfnamefont {M.~A.~F.}\ \bibnamefont {{de Souza}}}, \bibinfo {author}
  {\bibfnamefont {P.}~\bibnamefont {{Di Marcantonio}}}, \bibinfo {author}
  {\bibfnamefont {I.}~\bibnamefont {{Di Varano}}}, \bibinfo {author}
  {\bibfnamefont {V.}~\bibnamefont {{D'Odorico}}}, \bibinfo {author}
  {\bibfnamefont {H.}~\bibnamefont {{Drass}}}, \bibinfo {author} {\bibfnamefont
  {P.}~\bibnamefont {{Figueira}}}, \bibinfo {author} {\bibfnamefont {A.~B.}\
  \bibnamefont {{Fragoso}}}, \bibinfo {author} {\bibfnamefont {J.~P.~U.}\
  \bibnamefont {{Fynbo}}}, \bibinfo {author} {\bibfnamefont {M.}~\bibnamefont
  {{Genoni}}}, \bibinfo {author} {\bibfnamefont {J.~I.}\ \bibnamefont
  {{Gonz{\'a}lez Hern{\'a}ndez}}}, \bibinfo {author} {\bibfnamefont
  {M.}~\bibnamefont {{Haehnelt}}}, \bibinfo {author} {\bibfnamefont
  {I.}~\bibnamefont {{Hughes}}}, \bibinfo {author} {\bibfnamefont
  {P.}~\bibnamefont {{Huke}}}, \bibinfo {author} {\bibfnamefont
  {H.}~\bibnamefont {{Kjeldsen}}}, \bibinfo {author} {\bibfnamefont {A.~J.}\
  \bibnamefont {{Korn}}}, \bibinfo {author} {\bibfnamefont {M.}~\bibnamefont
  {{Land oni}}}, \bibinfo {author} {\bibfnamefont {J.}~\bibnamefont {{Liske}}},
  \bibinfo {author} {\bibfnamefont {C.}~\bibnamefont {{Lovis}}}, \bibinfo
  {author} {\bibfnamefont {R.}~\bibnamefont {{Maiolino}}}, \bibinfo {author}
  {\bibfnamefont {T.}~\bibnamefont {{Marquart}}}, \bibinfo {author}
  {\bibfnamefont {C.~J.~A.~P.}\ \bibnamefont {{Martins}}}, \bibinfo {author}
  {\bibfnamefont {E.}~\bibnamefont {{Mason}}}, \bibinfo {author} {\bibfnamefont
  {M.~A.}\ \bibnamefont {{Monteiro}}}, \bibinfo {author} {\bibfnamefont
  {T.}~\bibnamefont {{Morris}}}, \bibinfo {author} {\bibfnamefont
  {G.}~\bibnamefont {{Murray}}}, \bibinfo {author} {\bibfnamefont
  {A.}~\bibnamefont {{Niedzielski}}}, \bibinfo {author} {\bibfnamefont
  {E.}~\bibnamefont {{Oliva}}}, \bibinfo {author} {\bibfnamefont
  {L.}~\bibnamefont {{Origlia}}}, \bibinfo {author} {\bibfnamefont
  {E.}~\bibnamefont {{Pall{\'e}}}}, \bibinfo {author} {\bibfnamefont
  {P.}~\bibnamefont {{Parr-Burman}}}, \bibinfo {author} {\bibfnamefont {V.~C.}\
  \bibnamefont {{Parro}}}, \bibinfo {author} {\bibfnamefont {F.}~\bibnamefont
  {{Pepe}}}, \bibinfo {author} {\bibfnamefont {N.}~\bibnamefont {{Piskunov}}},
  \bibinfo {author} {\bibfnamefont {J.~L.}\ \bibnamefont {{Rasilla}}}, \bibinfo
  {author} {\bibfnamefont {P.}~\bibnamefont {{Rees}}}, \bibinfo {author}
  {\bibfnamefont {R.}~\bibnamefont {{Rebolo}}}, \bibinfo {author}
  {\bibfnamefont {M.}~\bibnamefont {{Riva}}}, \bibinfo {author} {\bibfnamefont
  {S.}~\bibnamefont {{Rousseau}}}, \bibinfo {author} {\bibfnamefont
  {N.}~\bibnamefont {{Sanna}}}, \bibinfo {author} {\bibfnamefont {N.~C.}\
  \bibnamefont {{Santos}}}, \bibinfo {author} {\bibfnamefont {T.~C.}\
  \bibnamefont {{Shen}}}, \bibinfo {author} {\bibfnamefont {F.}~\bibnamefont
  {{Sortino}}}, \bibinfo {author} {\bibfnamefont {D.}~\bibnamefont
  {{Sosnowska}}}, \bibinfo {author} {\bibfnamefont {S.}~\bibnamefont
  {{Sousa}}}, \bibinfo {author} {\bibfnamefont {E.}~\bibnamefont {{Stempels}}},
  \bibinfo {author} {\bibfnamefont {K.}~\bibnamefont {{Strassmeier}}}, \bibinfo
  {author} {\bibfnamefont {F.}~\bibnamefont {{Tenegi}}}, \bibinfo {author}
  {\bibfnamefont {A.}~\bibnamefont {{Tozzi}}}, \bibinfo {author} {\bibfnamefont
  {S.}~\bibnamefont {{Udry}}}, \bibinfo {author} {\bibfnamefont
  {L.}~\bibnamefont {{Valenziano}}}, \bibinfo {author} {\bibfnamefont
  {L.}~\bibnamefont {{Vanzi}}}, \bibinfo {author} {\bibfnamefont
  {M.}~\bibnamefont {{Weber}}}, \bibinfo {author} {\bibfnamefont
  {M.}~\bibnamefont {{Woche}}}, \bibinfo {author} {\bibfnamefont
  {M.}~\bibnamefont {{Xompero}}}, \ and\ \bibinfo {author} {\bibfnamefont
  {E.}~\bibnamefont {{Zackrisson}}},\ }in\ \href {\doibase 10.1117/12.2311664}
  {\emph {\bibinfo {booktitle} {Society of Photo-Optical Instrumentation
  Engineers (SPIE) Conference Series}}},\ Vol.\ \bibinfo {volume} {10702}\
  (\bibinfo {year} {2018})\ p.\ \bibinfo {pages} {107021Y}\BibitemShut
  {NoStop}%
\bibitem [{\citenamefont {{Murphy}}\ \emph
  {et~al.}(2016{\natexlab{a}})\citenamefont {{Murphy}}, \citenamefont
  {{Malec}},\ and\ \citenamefont {{Prochaska}}}]{Murphy2016}%
  \BibitemOpen
  \bibfield  {author} {\bibinfo {author} {\bibfnamefont {M.~T.}\ \bibnamefont
  {{Murphy}}}, \bibinfo {author} {\bibfnamefont {A.~L.}\ \bibnamefont
  {{Malec}}}, \ and\ \bibinfo {author} {\bibfnamefont {J.~X.}\ \bibnamefont
  {{Prochaska}}},\ }\href {\doibase 10.1093/mnras/stw1482} {\bibfield
  {journal} {\bibinfo  {journal} {Monthly Notices Royal Astronomial Society}\
  }\textbf {\bibinfo {volume} {461}},\ \bibinfo {pages} {2461} (\bibinfo {year}
  {2016}{\natexlab{a}})},\ \Eprint {http://arxiv.org/abs/1606.06293}
  {arXiv:1606.06293} \BibitemShut {NoStop}%
\bibitem [{\citenamefont {{Songaila}}\ and\ \citenamefont
  {{Cowie}}(2014)}]{Songaila2014}%
  \BibitemOpen
  \bibfield  {author} {\bibinfo {author} {\bibfnamefont {A.}~\bibnamefont
  {{Songaila}}}\ and\ \bibinfo {author} {\bibfnamefont {L.~L.}\ \bibnamefont
  {{Cowie}}},\ }\href {\doibase 10.1088/0004-637X/793/2/103} {\bibfield
  {journal} {\bibinfo  {journal} {\apj}\ }\textbf {\bibinfo {volume} {793}},\
  \bibinfo {eid} {103} (\bibinfo {year} {2014})},\ \Eprint
  {http://arxiv.org/abs/1406.3628} {arXiv:1406.3628} \BibitemShut {NoStop}%
\bibitem [{\citenamefont {{Kotu{\v s}}}\ \emph {et~al.}(2017)\citenamefont
  {{Kotu{\v s}}}, \citenamefont {{Murphy}},\ and\ \citenamefont
  {{Carswell}}}]{Kotus2017}%
  \BibitemOpen
  \bibfield  {author} {\bibinfo {author} {\bibfnamefont {S.~M.}\ \bibnamefont
  {{Kotu{\v s}}}}, \bibinfo {author} {\bibfnamefont {M.~T.}\ \bibnamefont
  {{Murphy}}}, \ and\ \bibinfo {author} {\bibfnamefont {R.~F.}\ \bibnamefont
  {{Carswell}}},\ }\href {\doibase 10.1093/mnras/stw2543} {\bibfield  {journal}
  {\bibinfo  {journal} {Monthly Notices Royal Astronomial Society}\ }\textbf
  {\bibinfo {volume} {464}},\ \bibinfo {pages} {3679} (\bibinfo {year}
  {2017})},\ \Eprint {http://arxiv.org/abs/1609.03860} {arXiv:1609.03860}
  \BibitemShut {NoStop}%
\bibitem [{\citenamefont {{Agafonova}}\ \emph {et~al.}(2011)\citenamefont
  {{Agafonova}}, \citenamefont {{Molaro}}, \citenamefont {{Levshakov}},\ and\
  \citenamefont {{Hou}}}]{Agafonova2011}%
  \BibitemOpen
  \bibfield  {author} {\bibinfo {author} {\bibfnamefont {I.~I.}\ \bibnamefont
  {{Agafonova}}}, \bibinfo {author} {\bibfnamefont {P.}~\bibnamefont
  {{Molaro}}}, \bibinfo {author} {\bibfnamefont {S.~A.}\ \bibnamefont
  {{Levshakov}}}, \ and\ \bibinfo {author} {\bibfnamefont {J.~L.}\ \bibnamefont
  {{Hou}}},\ }\href {\doibase 10.1051/0004-6361/201016194} {\bibfield
  {journal} {\bibinfo  {journal} {Astronomy and Astrophysics}\ }\textbf
  {\bibinfo {volume} {529}},\ \bibinfo {eid} {A28} (\bibinfo {year} {2011})},\
  \Eprint {http://arxiv.org/abs/1102.2967} {arXiv:1102.2967} \BibitemShut
  {NoStop}%
\bibitem [{\citenamefont {{Bainbridge}}\ and\ \citenamefont
  {{Webb}}(2017)}]{Bainbridge2017}%
  \BibitemOpen
  \bibfield  {author} {\bibinfo {author} {\bibfnamefont {M.~B.}\ \bibnamefont
  {{Bainbridge}}}\ and\ \bibinfo {author} {\bibfnamefont {J.~K.}\ \bibnamefont
  {{Webb}}},\ }\href {\doibase 10.1093/mnras/stx179} {\bibfield  {journal}
  {\bibinfo  {journal} {Monthly Notices Royal Astronomical Society}\ }\textbf
  {\bibinfo {volume} {468}},\ \bibinfo {pages} {1639} (\bibinfo {year}
  {2017})},\ \Eprint {http://arxiv.org/abs/1606.07393} {arXiv:1606.07393
  [astro-ph.IM]} \BibitemShut {NoStop}%
\bibitem [{\citenamefont {{Leite}}\ \emph {et~al.}(2016)\citenamefont
  {{Leite}}, \citenamefont {{Martins}}, \citenamefont {{Molaro}}, \citenamefont
  {{Corre}},\ and\ \citenamefont {{Cristiani}}}]{Leite2016}%
  \BibitemOpen
  \bibfield  {author} {\bibinfo {author} {\bibfnamefont {A.~C.~O.}\
  \bibnamefont {{Leite}}}, \bibinfo {author} {\bibfnamefont {C.~J.~A.~P.}\
  \bibnamefont {{Martins}}}, \bibinfo {author} {\bibfnamefont {P.}~\bibnamefont
  {{Molaro}}}, \bibinfo {author} {\bibfnamefont {D.}~\bibnamefont {{Corre}}}, \
  and\ \bibinfo {author} {\bibfnamefont {S.}~\bibnamefont {{Cristiani}}},\
  }\href {\doibase 10.1103/PhysRevD.94.123512} {\bibfield  {journal} {\bibinfo
  {journal} {\prd}\ }\textbf {\bibinfo {volume} {94}},\ \bibinfo {eid} {123512}
  (\bibinfo {year} {2016})},\ \Eprint {http://arxiv.org/abs/1612.05284}
  {arXiv:1612.05284 [astro-ph.CO]} \BibitemShut {NoStop}%
\bibitem [{\citenamefont {{Murphy}}(2002)}]{Murphyphd}%
  \BibitemOpen
  \bibfield  {author} {\bibinfo {author} {\bibfnamefont {M.~T.}\ \bibnamefont
  {{Murphy}}},\ }\emph {\bibinfo {title} {{Probing variations in the
  fundamental constants with quasar absorption lines}}},\ \href@noop {} {Ph.D.
  thesis},\ \bibinfo  {school} {School of Physics, University of New South
  Wales} (\bibinfo {year} {2002})\BibitemShut {NoStop}%
\bibitem [{\citenamefont {{King}}(2012)}]{Kingphd}%
  \BibitemOpen
  \bibfield  {author} {\bibinfo {author} {\bibfnamefont {J.~A.}\ \bibnamefont
  {{King}}},\ }\emph {\bibinfo {title} {{Searching for variations in the
  fine-structure constant and the proton-to-electron mass ratio using quasar
  absorption lines}}},\ \href@noop {} {Ph.D. thesis},\ \bibinfo  {school}
  {School of Physics, University of New South Wales} (\bibinfo {year}
  {2012})\BibitemShut {NoStop}%
\bibitem [{\citenamefont {{Molaro}}\ \emph {et~al.}(2008)\citenamefont
  {{Molaro}}, \citenamefont {{Reimers}}, \citenamefont {{Agafonova}},\ and\
  \citenamefont {{Levshakov}}}]{Molaro2008}%
  \BibitemOpen
  \bibfield  {author} {\bibinfo {author} {\bibfnamefont {P.}~\bibnamefont
  {{Molaro}}}, \bibinfo {author} {\bibfnamefont {D.}~\bibnamefont {{Reimers}}},
  \bibinfo {author} {\bibfnamefont {I.~I.}\ \bibnamefont {{Agafonova}}}, \ and\
  \bibinfo {author} {\bibfnamefont {S.~A.}\ \bibnamefont {{Levshakov}}},\
  }\href {\doibase 10.1140/epjst/e2008-00818-4} {\bibfield  {journal} {\bibinfo
   {journal} {European Physical Journal Special Topics}\ }\textbf {\bibinfo
  {volume} {163}},\ \bibinfo {pages} {173} (\bibinfo {year} {2008})},\ \Eprint
  {http://arxiv.org/abs/0712.4380} {arXiv:0712.4380} \BibitemShut {NoStop}%
\bibitem [{\citenamefont {{Murphy}}\ \emph
  {et~al.}(2016{\natexlab{b}})\citenamefont {{Murphy}}, \citenamefont
  {{Malec}},\ and\ \citenamefont {{Prochaska}}}]{Murphy16}%
  \BibitemOpen
  \bibfield  {author} {\bibinfo {author} {\bibfnamefont {M.~T.}\ \bibnamefont
  {{Murphy}}}, \bibinfo {author} {\bibfnamefont {A.~L.}\ \bibnamefont
  {{Malec}}}, \ and\ \bibinfo {author} {\bibfnamefont {J.~X.}\ \bibnamefont
  {{Prochaska}}},\ }\href {\doibase 10.1093/mnras/stw1482} {\bibfield
  {journal} {\bibinfo  {journal} {Monthly Notices Royal Astronomical Society}\
  }\textbf {\bibinfo {volume} {461}},\ \bibinfo {pages} {2461} (\bibinfo {year}
  {2016}{\natexlab{b}})},\ \Eprint {http://arxiv.org/abs/1606.06293}
  {arXiv:1606.06293} \BibitemShut {NoStop}%
\bibitem [{\citenamefont {{Avgoustidis}}\ \emph {et~al.}(2009)\citenamefont
  {{Avgoustidis}}, \citenamefont {{Verde}},\ and\ \citenamefont
  {{Jimenez}}}]{AVJ2009}%
  \BibitemOpen
  \bibfield  {author} {\bibinfo {author} {\bibfnamefont {A.}~\bibnamefont
  {{Avgoustidis}}}, \bibinfo {author} {\bibfnamefont {L.}~\bibnamefont
  {{Verde}}}, \ and\ \bibinfo {author} {\bibfnamefont {R.}~\bibnamefont
  {{Jimenez}}},\ }\href {\doibase 10.1088/1475-7516/2009/06/012} {\bibfield
  {journal} {\bibinfo  {journal} {Journal of Cosmology and Astro-Particle
  Physics}\ }\textbf {\bibinfo {volume} {2009}},\ \bibinfo {eid} {012}
  (\bibinfo {year} {2009})},\ \Eprint {http://arxiv.org/abs/0902.2006}
  {arXiv:0902.2006 [astro-ph.CO]} \BibitemShut {NoStop}%
\bibitem [{\citenamefont {{Lv}}\ and\ \citenamefont {{Xia}}(2016)}]{Lv:2016}%
  \BibitemOpen
  \bibfield  {author} {\bibinfo {author} {\bibfnamefont {M.-Z.}\ \bibnamefont
  {{Lv}}}\ and\ \bibinfo {author} {\bibfnamefont {J.-Q.}\ \bibnamefont
  {{Xia}}},\ }\href {\doibase 10.1016/j.dark.2016.06.003} {\bibfield  {journal}
  {\bibinfo  {journal} {Physics of the Dark Universe}\ }\textbf {\bibinfo
  {volume} {13}},\ \bibinfo {pages} {139} (\bibinfo {year} {2016})},\ \Eprint
  {http://arxiv.org/abs/1606.08102} {arXiv:1606.08102 [astro-ph.CO]}
  \BibitemShut {NoStop}%
\end{thebibliography}%

\end{document}